\pdfoutput=1

\documentclass[fleqn,usenatbib]{mnras}

\usepackage{newtxtext,newtxmath}
\usepackage[T1]{fontenc}

\DeclareRobustCommand{\VAN}[3]{#2}
\let\VANthebibliography\thebibliography
\def\thebibliography{\DeclareRobustCommand{\VAN}[3]{##3}\VANthebibliography}

\usepackage{graphicx}	% Including figure files
\usepackage{amsmath}	% Advanced maths commands
\usepackage{amssymb}	% Extra maths symbols

\newcommand{\aref}[1]{\hyperref[#1]{Appendix~\ref{#1}}}

\newcommand{\epsff}{\epsilon_{\rm ff}}
\newcommand{\alphad}{\alpha_{\rm d}}
\newcommand{\alphag}{\alpha_{\rm g}}
\newcommand{\sd}{s_{\rm d}}
\newcommand{\sg}{s_{\rm g}}
\newcommand{\avir}{\alpha_{\rm vir,0}}
\newcommand{\mach}{\mathcal{M}}
\newcommand{\casemach}{\mathcal{M}\approx}
\newcommand{\alphavir}{\alpha_{\rm vir}}

\usepackage{xcolor}
\usepackage[normalem]{ulem}

\definecolor{darkgreen}{rgb}{0.13, 0.55, 0.13}
\definecolor{rubinered}{rgb}{0.82, 0.0, 0.34}
\definecolor{orange(ryb)}{rgb}{0.98, 0.4, 0.01}
\definecolor{scut}{rgb}{0.58, 0.22, 0.51}

\defcitealias{FK12}{FK12}

\title[
Supersonic self-gravitating turbulence]{
The density structure of supersonic self-gravitating turbulence
}

\author[S. Khullar et al.]{Shivan Khullar,$^{1,2}$\thanks{E-mail: khullar@astro.utoronto.ca (SK)}
Christoph Federrath,$^{3,4}$
Mark R. Krumholz, $^{3,4}$
\newauthor
Christopher D.\ Matzner$^{1}$
\\
$^{1}$David A. Dunlap Dept.\ of Astronomy \& Astrophysics, University of Toronto, 50 St. George St., Toronto, ON M5S 3H4, Canada\\
$^{2}$Canadian Institute for Theoretical Astrophysics, University of Toronto, 60 St. George St., Toronto, ON M5S 3H8, Canada\\
$^{3}$Research School of Astronomy and Astrophysics, Australian National University, Cotter Rd., Weston ACT 2611, Australia\\
$^{4}$ARC Centre of Excellence for Astronomy in Three Dimensions (ASTRO-3D), Canberra ACT 2601, Australia
}

\date{Accepted XXX. Received YYY; in original form ZZZ}

\pubyear{2021}

\begin{document}
\label{firstpage}
\pagerange{\pageref{firstpage}--\pageref{lastpage}}
\maketitle

\begin{abstract}
We conduct numerical experiments to determine the density probability distribution function (PDF) produced in supersonic, isothermal, self-gravitating turbulence of the sort that is ubiquitous in star-forming molecular clouds. Our experiments cover a wide range of turbulent Mach number and virial parameter, allowing us for the first time to determine how the PDF responds as these parameters vary, and we introduce a new diagnostic, the dimensionless star formation efficiency versus density ($\epsff(s)$) curve, which provides a sensitive diagnostic of the PDF shape and dynamics. We show that the PDF follows a universal functional form consisting of a log-normal at low density with two distinct power law tails at higher density; the first of these represents the onset of self-gravitation, and the second reflects the onset of rotational support. Once the star formation efficiency reaches a few percent, the PDF becomes statistically steady, with no evidence for secular time-evolution at star formation efficiencies from about five to 20 percent.  We show that both the Mach number and the virial parameter influence the characteristic densities at which the log-normal gives way to the first power-law, and the first to the second, and we extend (for the former) and develop (for the latter) simple theoretical models for the relationship between these density thresholds and the global properties of the turbulent medium.
\end{abstract}


%%%%%%%%%%%%%%%%%%%%%%%%%%%%%%%%%%%%%%%%%%%%%%%%%%

%%%%%%%%%%%%%%%%% BODY OF PAPER %%%%%%%%%%%%%%%%%%

\section{Introduction}
\label{sec:Introduction}

The interplay of turbulence and gravity lies at the heart of many astrophysical problems. Over the last few decades, there has been tremendous progress toward understanding astrophysical turbulence in a statistical sense (e.g., \citealt{Vazquez-Semadeni1994, Padoan1997, NordlundPadoan1999, Elmegreen2002b,  Kritsuk2007}). The density probability distribution function (PDF) has shown particular promise as a tool to study turbulent gas flows in the presence of gravity (e.g. \citealt{Ballesteros-Paredes1998, Elmegreen2002a, KM05, DibEtAl2007, FederrathKlessenSchmidt2008, FederrathBanerjee2015, Nolan2015, Murray2015, FederrathEtAl2016CMZ, Burkhart2016, Pan2018, Pan2019, Jaupart2020}). 
The density PDF serves as an important input to analytical theories of turbulence-regulated star formation, which link estimates of the star formation rate or the initial mass function of stars directly to the PDF (see e.g., \citealt{Padoan2002, Padoan2004, KM05, Padoan2007, Hennebelle2008, Hennebelle2009, HC11, PN11,FK12,Hopkins2013b}).  It is also closely related to the directly-observable column density distribution function, the N-PDF (\citealt{Brunt2010a, Brunt2010b, Ginsburg2013, Kainulainen2014}). For these reasons, it is critical to understand the physical processes shaping the density PDF. 

A number of studies have shown that, for supersonic isothermal turbulence, the
density PDF assumes a log-normal (LN) distribution, at least for low Mach numbers \citep{Vazquez-Semadeni1994, Padoan1997, Ostriker2001, Kritsuk2007, Lemaster2008, FederrathKlessenSchmidt2008, Federrath2010, Konstandin2012, Molina2012, Nolan2015}. This would be expected from an application of the central limit theorem: the build up of density ($\rho$) fluctuations (such as shocks) should be random and multiplicative in nature, or equivalently, for the variable $s =  \ln(\rho/\rho_0)$, this is a random additive process which will result in a Gaussian distribution (LN in $\rho$: \citealt{Pope1993,Passot1998}). 
However, some numerical simulations have shown deviations from the LN distribution as well (\citealt{Federrath2008, Schmidt2009, Federrath2010, Price2010, Konstandin2012, FK13, Pan2019}). \citet{Hopkins2013} argues that these appear because the density fluctuations are not uncorrelated and thus the central limit theorem is not fully applicable. 
Despite these caveats, the LN distribution remains a good approximation for turbulent supersonic media, and is consistent with observations of molecular clouds
(\citealt{Berkhuijsen2008, Hill2008, Burkhart2010, Burkhart2012, Maier2016, Federrath2016, Maier2017, Chen2018}; however, for a contrary view, see \citealt{Lombardi2015, Alves2017} who suggest that the distribution follows a power law modified at low densities by the finite size of a given map).

Real molecular clouds are self-gravitating in addition to being magnetized and turbulent, and 
virtually all numerical studies that include gravity find that it modifies the density PDF  (e.g. \citealt{Ballesteros-Paredes2011, Kritsuk2011, Collins2012, FK13,Girichidis2014, Burkhart2015a}).  Self-gravity causes the turbulent (LN) distribution to develop a power-law (PL) tail at high densities.  An analogous tail is observed in the N-PDF of several observed molecular clouds (e.g.,\citealt{Schneider2013, Schneider2015a, Schneider2015b, Pokhrel2016, Dib2020}).  
This PL tail forms when gravitational forces become important, such as in the process of collapse. \citet{Kritsuk2011} invoke models based on the collapse of singular isothermal spheres (\citealt{Penston1969a, Penston1969b, Larson1969, Shu1977, Hunter1977, Whitworth1985}) to explain the exponent of the PL tail. 
\citet{Collins2012} find that the inclusion of self-gravity leads to the division of the density PDF into two distinct phases, a low-density turbulent cloud and high-density, self-gravitating cores.
\citet{Girichidis2014} develop an analytical description for how this density PDF evolves during the free-fall collapse of a uniform density sphere. These authors find that the PL tail of the (volume-weighted) density PDF asymptotes to a slope $-1.54$. They argue that physical processes such as driven converging flows can accelerate collapse and flatten the slope of the PL tail. 
\citet{Jaupart2020} build on the analytical results of \citet{Pan2018, Pan2019} and develop a theory for the shape and evolution of the density PDF from the Navier-Stokes and Poisson equations. They determine a density threshold above which gravity would dominate the dynamics of turbulence, which depends on the cloud parameters such as the Mach number and virial parameter. In their theory, PL tails develop naturally over time-scales of a few mean free-fall times.

High-resolution simulations also suggest the existence of a second PL tail at higher densities, which is distinguished from the first tail by having a different slope \citep{Kritsuk2011, Collins2011, Collins2012, MurrayChangMurray2017, Veltchev2019a, Marinkova2020}. However, the origin of this feature remains disputed. \citet{Kritsuk2011} and \citet{MurrayChangMurray2017} suggest that it is due to pile up of mass caused by the formation of centrifugally-supported discs, but \citeauthor{Kritsuk2011} also speculate that it could be a numerical artefact created when simulations violate the \citet{Truelove1997} resolution condition at the highest densities. \citet{Jaupart2020} argue that a second PL tail, with a characteristic slope of 1.5, is a signature of regions in free-fall collapse. \citet{Donkov2020} model clouds as self-gravitating spheres, and treat turbulent support as a modified equation of state (following \citealt{Donkov2018, Donkov2019}). In their model, a second PL tail emerges as a result of the transition between predominantly turbulent support at larger radii and predominantly thermal support at smaller radii. There is also some observational evidence for second PL tails in N-PDFs, mostly from analysis of \textit{Herschel}-derived dust column density maps \citep{Tremblin2014, Schneider2015, Pokhrel2016, Dib2020}. It is unclear, however, if these features are due to the same processes that drive the second PL features in simulated volume density PDFs.

Current uncertainties about the functional form of the density PDF, and its physical origin, provide the primary motivation for this paper. To test these, we perform a suite of simulations of isothermal gravo-turbulent fluids in which we systematically vary the dimensionless parameters -- the Mach number, $\mach$, and the virial parameter, $\avir$ -- that describe the gas. We characterise the variation of the PDF with these parameters and over time within a single simulation, and present analytic models to interpret these results. In the process, we make extensive use of the intimate links between density PDFs and the star formation rate per free-fall time, $\epsilon_{\rm ff}$, which is one of the central quantities that any successful theory of star formation should be able to predict (e.g.~ \citealt{KM05,FK12,FK13, SFK15, Vutisalchavakul2016, Heyer16a, Leroy17a, Grisdale2019}). \citet{Khullar19} show that the variation of this quantity with density, $\epsff(\rho)$, is a powerful diagnostic that can be used to identify changes in behaviour at characteristic densities, often referred to as star formation thresholds, because it characterizes the rate at which gravity drives the evolution from low to high densities. We extend this analysis to show that $\epsilon(\rho)$ can also be used to understand the origin and variation of the density PDF, and that the two functions can be regarded as two sides of the same coin.

The rest of the paper is organized as follows: in \autoref{sec:Methodology}, we describe our numerical simulations and our method for extracting PDF information from them. In \autoref{sec:Results}, we discuss the physical processes controlling the density PDF and look at the variation of the PDF parameters with the simulation parameters. In \autoref{sec:disc-calculation}, we present models for the dependence of the PDF on Mach number and virial parameter. We summarize our conclusions in \autoref{sec:Conclusions}. 

\section{Simulations and Methods}
\label{sec:Methodology}

\subsection{Simulations}

\begin{table*}
	\centering
	\caption{Summary of the simulation suite. Column 1: Simulation name. Columns 2-4: Virial Parameter as defined by equation \ref{eq:vir-analytic}, (3D) Velocity dispersion, Mach number $\mach$. Columns 5-7: Box size, Effective grid size (resolution), Effective minimum cell size. Columns 8-10: Total mass, Mean density, Sink particle threshold density. Columns 11-12: Autocorrelation Time, Free-fall time at the mean density.}
	\begin{tabular}{lcccccccccccr} 
		\hline
		Name & $\alpha_{\rm vir,o}$ & $\sigma_V$ & $\mathcal{M}$ &  $L$ & $N_{\rm res}$  & $\Delta x$ & $M$ & $\rho_0$ & $\rho_{\rm sink}$ & $T_{\rm auto}$ & $t_{\rm ff}$ \\
		& & [km/s] & & [pc] & & [AU] & [M$_{\odot}$] & [g cm$^{-3}$] & [g cm$^{-3}$]  &  [Myr] &  [Myr] \\
		(1) & (2) & (3) & (4) & (5) & (6) & (7) & (8) & (9) & (10) & (11) & (12)\\
		\hline
		V0.5M2.5 & 0.5 & 0.5 & 2.5 & 2 & 2048$^3$ & 201.45 & 193.86 & 1.64 $\times$ 10$^{-21}$ & 8.29 $\times$ 10$^{-17}$ & 1.957 & 1.644\\
		V1M2.5 & 1 & 0.5 & 2.5 & 2 & 2048$^3$ & 201.45 & 96.93 & 0.82 $\times$ 10$^{-21}$ & 8.29 $\times$ 10$^{-17}$ & 1.957 & 2.326\\ \hline
		V0.5M5 & 0.5 & 1.0 & 5 & 2 & 2048$^3$ & 201.45 & 775.44 & 6.56 $\times$ 10$^{-21}$ & 8.29 $\times$ 10$^{-17}$ & 0.978 & 0.822\\
		V1M5 & 1 & 1.0 & 5 & 2 & 2048$^3$ & 201.45 & 387.72 & 3.28 $\times$ 10$^{-21}$ & 8.29 $\times$ 10$^{-17}$ & 0.978 & 1.163\\
		V2M5 & 2 & 1.0 & 5 & 2 & 2048$^3$ & 201.45 & 193.86 & 1.64 $\times$ 10$^{-21}$ & 8.29 $\times$ 10$^{-17}$ & 0.978 & 1.645\\ \hline
		V0.5M10 & 0.5 & 2.0 & 10 & 2 & 2048$^3$ & 201.45 & 3101.75 & 2.62 $\times$ 10$^{-20}$ & 8.29 $\times$ 10$^{-17}$ & 0.489 & 0.411 \\
		V1M10 & 1 & 2.0 & 10 & 2 & 8192$^3$ & 50.35 & 1550.87 & 1.31 $\times$ 10$^{-20}$ & 1.33 $\times$ 10$^{-15}$ & 0.489& 0.582 \\
		V2M10 & 2 & 2.0 & 10 & 2 & 2048$^3$ & 201.45 & 775.44 & 6.56 $\times$ 10$^{-21}$ & 8.29 $\times$ 10$^{-17}$ & 0.489 & 0.822\\ \hline
        V2M5res256 & 2 & 1.0 & 5 & 2 & 256$^3$ & 1611.61 & 193.86 & 1.64 $\times$ 10$^{-21}$ & 1.30 $\times$ 10$^{-18}$ & 0.978 & 1.645\\ 
        V2M5res512 & 2 & 1.0 & 5 & 2 & 512$^3$ & 805.80 & 193.86 & 1.64 $\times$ 10$^{-21}$ & 5.18 $\times$ 10$^{-18}$ & 0.978 & 1.645\\
        V2M5res1024 & 2 & 1.0 & 5 & 2 & 1024$^3$ & 402.90 & 193.86 & 1.64 $\times$ 10$^{-21}$ & 2.08 $\times$ 10$^{-17}$ & 0.978 & 1.645\\ 
        V1M10res2048 & 1 & 2.0 & 10 & 2 & 2048$^3$ & 201.45 & 1550.87 & 1.31 $\times$ 10$^{-20}$ & 8.29 $\times$ 10$^{-17}$ & 0.489& 0.582 \\
        V2M5res4096 & 2 & 1.0 & 5 & 2 & 4096$^3$ & 100.73 & 193.86 & 1.64 $\times$ 10$^{-21}$ & 3.32 $\times$ 10$^{-16}$ & 0.978 & 1.645\\ 
		\hline
		
	\end{tabular}
	\label{tab:Sim-params}
\end{table*}

We perform a suite of 13 simulations using the Adaptive Mesh Refinement (AMR, \citealt{Berger1989}) code FLASH \citep{Fryxell2000, Dubey2008}. We use a setup similar to that of \citet[hereafter, \citetalias{FK12}]{FK12}, but with higher resolution, and encompassing a range of dimensionless parameters necessary to study the variation of the density PDF with them. We solve the following equations of hydrodynamics and self-gravity in three dimensions:

\begin{align}
    \label{eq:ContinuityHD}
    & \frac{\partial \rho}{\partial t} + \mathbf{\nabla}\cdot(\rho \mathbf{v}) = 0, \\
    & \rho \frac{\partial \mathbf{v}}{\partial t} + \rho (\mathbf{v}\cdot\mathbf{\nabla})\mathbf{v} = - \rho(\mathbf{g} + \mathbf{F_{\rm turb}}) - \mathbf{\nabla}(P_{\rm th}),
    \label{eq:momentum}
    \\
    & \mathbf{g} = -\mathbf{\nabla}\phi_{\rm gas} + \mathbf{g_{\rm sink}},\\
    & \nabla^2 \phi_{\rm gas} = 4\pi G \rho,
\end{align}
%\begin{equation}
%\label{eq:EnergyHD}
 %   \frac{\partial E}{\partial t} + \nabla\cdot (E + P_{\rm th}\mathbf{v}) =  %\rho\mathbf{v}\cdot(\mathbf{g} + \mathbf{F_{\rm turb})},
%\end{equation}
where $\rho$ and $\mathbf{v}$ denote the gas density and velocity respectively, and $\mathbf{g}_{\rm sink}$ is the gravity of the sink particles (\S\,\ref{sec:Sinks}).
%The term $\mathbf{g}$ includes the contribution from the self-gravity of the gas and acceleration due to gravity of the sink particles (see section \ref{sec:Sinks} below). 
%The total energy density $E = 1/2 \rho |\mathbf{v}|^2 + \rho \epsilon_{\rm int}$ (the sum of the thermal and internal energy densities). 
Finally, to close the hydrodynamics equations we use an isothermal equation of state $P_{\rm th} = c_s^2 \rho$ with uniform sound speed $c_s = ({k_B T/\mu m_{\rm H}})^{1/2}$.  While each simulation can be scaled to different physical parameters, we nominally set $c_s= 0.2$ km s$^{-1}$, corresponding to gas temperature  $T\approx 11$ K and mean molecular weight  $\mu=2.3$. The isothermal approximation is reasonable for molecular cloud environments \citep{KrumholzSFBook2015} up to a characteristic critical density $\sim 10^{14}$ cm$^{-3}$ \citep{Masunaga2000}. Although our simulations do exceed this density at times, we maintain the isothermal approximation in order to focus on the hydrodynamics of the density distribution, rather than the thermal physics that determines fragmentation.  

Our simulation domain is a periodic box of size $L = 2$~pc on each side. We discretise the box with maximum effective resolutions of $N_{\rm res} = 512^3$--$8192^3$, with a base grid of 512$^3$ with up to 5 levels of AMR and a refinement criterion that kicks in when the local Jeans length, $\lambda_J = (\pi c_s^2/G\rho)^{1/2}$, falls below 16~grid cells. This refinement criterion ensures that the local Jeans length is well-resolved
%by at least 16~grid cells
at any point in space and time, and that the turbulence can be reasonably well resolved on the Jeans scale. %(strictly-speaking, 30 grid cells are required to fully resolve the solenoidal energy component of turbulence on the Jeans scale; see Federrath et al. 2011). 
We use a positive-definite Riemann solver \citep{Waagan2011} for solving the system of hydrodynamic equations and a multigrid Poisson solver \citep{Ricker2008} for the self-gravity of the gas with periodic boundary conditions. 

We caution readers that our simulations do not include magnetic fields or stellar feedback, both of which affect the star formation efficiency and thus likely the density PDF as well \citep[e.g.,][]{MatznerMcKee2000,Krumholz12b, Myers2014, F15, Schneider2015, Lin2016, Mocz2017, Grudic2018, C18}. However, including these effects would dramatically expand the parameter space, and in particular including feedback effects would likely make the results dependent on the details of the subgrid feedback recipe adopted. For this reason, purely hydrodynamic simulations, although less realistic, offer the advantage of a clean, more easily interpreted experiment. They are also valuable in that they form a baseline case against which magnetic and feedback effects can be identified in future work.
%, and are run at finite resolution (up to a maximum effective resolution of 8192$^3$ cells) within periodic boxes. 
%All of these are likely to effect the outcome of the density PDF and the star formation efficiency, as indicated by \citealt{F15, Schneider2015, Lin2016, Mocz2017, Grudic2018}. However, conclusions drawn from our numerical experiments can nevertheless be helpful.

%We briefly highlight the main features of our simulations relevant to this work below and refer the reader to FK12. 

%\mrknote{Move this paragraph to the end of section 2.1.1; it only makes sense once you describe the basic setup, i.e., you first drive turbulence to some steady state velocity dispersion. Here you introduce the symbol $\sigma_V$, but you haven't said anywhere what it means.}

%\footnote{$b=0.33$ for solenoidal driving (divergence-free), $b=1$ for compressive driving (curl-free) and $b=0.4$ for mixed driving.}, \textit{(iii)} the turbulent Mach number $\mathcal{M}$, and \textit{(iv)} 

\subsubsection{Turbulence Driving and Initial Conditions}
\label{sec:DrivingICs}

Our simulations run in two phases: we first drive the turbulence to statistical steady state with gravity turned off, and then turn on gravity and allow collapse. The procedure used to drive turbulence is similar to previous simulations (\citetalias{FK12}; \citealt{F15, Mathew2020}), and uses the basic method described in detail in \citet{Federrath2010}. We apply a stochastic Ornstein-Uhlenbeck process \citep{Eswaran1988} to construct an acceleration field $\mathbf{F}_{\rm turb}$, which appears as a source term in \autoref{eq:momentum}. $\mathbf{F}_{\rm turb}$ contains only large-scale modes, $1\leq k\leq3$ (where the wavenumber $k$ is in units of $2\pi/L$), and most of the power is injected at the $k = 2$ mode, corresponding to half the box size. Our turbulence driving module allows us to control the mixture of modes $\mathbf{F}_{\rm turb}$ excites, so we can drive with purely solenoidal modes ($\nabla \cdot \mathbf{F}_{\rm turb}=0$), purely compressive modes ($\nabla \times \mathbf{F}_{\rm turb}=0$), or a natural mixture of the two. We use natural mixtures for all the simulations presented here \citep{Federrath2010}.

%We shall denote this length scale as $\Tilde{L}=L/2$, hereafter. 

We initialize the simulation box with gas having uniform density $\rho_0$ and zero velocity, then drive turbulence for two turnover (or auto-correlation) times, $T_{\rm auto} = L/2\sigma_V$, to allow the turbulent cascade to build up. Here $\sigma_V$ is the steady-state velocity dispersion produced by our driving. We switch on self-gravity when $t_{\rm tot}=2T_{\rm auto}$ and denote this time as $t=0$. We continue to drive turbulence for all times during the evolution of the simulated clouds.

Our simulations can be characterized by three key dimensionless parameters: (a) the Mach number of the flow, $\mathcal{M} = \sigma_V/c_s$, (b) the virial ratio, $\alpha_\mathrm{vir}=2 E_{\rm kin}/|E_{\rm pot}|$, given as (under the assumption of a uniform, spherical gas cloud of radius $L/2$ in isolation),
\begin{equation}
    \label{eq:vir-analytic}
    \avir = \frac{5 \sigma_V^2 L}{6GM_c},    
\end{equation}
and (c) the turbulence driving parameter, $b$ (\citealt{Federrath2008, Federrath2010}). The symbols $E_{\rm kin}$, $E_{\rm pot}$, $\sigma_V$ and $M_c$ denote the kinetic energy, potential energy, three dimensional velocity dispersion, and mass of the gas cloud at $t=0$. Our choice of a natural mix of compressive and solenoidal modes corresponds to $b=0.4$ in all cases, so the simulations can be described by $\avir$ and $\mathcal{M}$ alone. We run simulations with $\avir=0.5$--$2$ and $\mathcal{M}\approx 2.5$--$10$. We summarize these and other key simulation parameters in \autoref{tab:Sim-params}.

\subsubsection{Sink Particles}
\label{sec:Sinks}

To model collapsing gas and stars in our simulations, we use sink particles (\citealt{Bate1997, Krumholz04, FederrathEtAl2010}). When the gas exceeds a specified density $\rho_{\rm sink}$ (at the maximum level of refinement), we create a spherical control volume with radius $r_{\rm sink}$, and perform additional checks on the collapsing region to determine if the gas in the control volume, for example, is bound and Jeans unstable. (For details, see \citealt{FederrathEtAl2010}.)
The sink particle threshold density is given by 
\begin{equation}
    \rho_{\rm sink} = \frac{\pi c_s^2}{G\lambda_{\rm J}^2},
\end{equation}
where $\lambda_{\rm J}$ is the Jeans length at $\rho =\rho_{\rm sink}$; we give numerical values of $\rho_{\rm sink}$ for our simulations in \autoref{tab:Sim-params}. The control volume radius $r_{\rm sink} = \lambda_{\rm J}/2$ at $\rho = \rho_{\rm sink}$ is set to 2.5 grid cell lengths at the maximum AMR level. This ensures that the Jeans length is resolved by at least 5 grid cells on the highest level of AMR, consistent with the \citet{Truelove1997} criterion to avoid artificial fragmentation. For more details about the implementation of sink particles, we refer the reader to \citet{FederrathEtAl2010,FederrathEtAl2014}.

\subsection{Density PDF fitting model} \label{sec:pdf_fitting}

A primary goal of this work is to study the shape of the volumetric density PDF, and its dependence on $\alpha_{\rm vir,0}$ and $\mathcal{M}$. For this purpose, it is helpful to describe the PDF in terms of a few simple parameters that we can extract by fitting to the distribution of cell densities found in our simulations (either in a single time snapshot or averaged over a number of snapshots). As we show below, and consistent with the conclusions of previous works (\autoref{sec:Introduction}), the PDFs in our simulations can be described reasonably well by a piecewise function consisting of a lognormal (LN) component at low density and one or two power-law (PL) components at high density.
We first introduce the single power-law form for the PDF since it has been used traditionally by previous works (see e.g., \citealt{Myers2015, BB17, Burkhart2018, Burkhart2019}).
In terms of the natural logarithm of the density, $s = \ln(\rho/\rho_0)$, where $\rho_0$ is the mean density in the computational volume, the single PL form, to which we shall henceforth refer as LN+PL, is given by
\begin{equation}
\label{eq:SPL}
p(s) =
\begin{cases}
\frac{N}{\sqrt{2 \pi \sigma_{\rm s}^2}} \exp\left[-\frac{(s-s_0)^2}{2 \sigma_{\rm s}^2}\right] & s<\sg, \\
N p_0 e^{-s \alpha_{\rm g}} & s\geq \sg,
\end{cases}
\end{equation}
where $\alphag$ is the slope of the power-law tail, $\sigma_{\rm s}$ is the width of the LN part, $s_0$ is the mean of the LN part and $N$ and $p_0$ are normalisation factors. This functional form contains 6~variables, $N, p_0, s_0, \sigma_{\rm s}, \sg, \alpha_{\rm g}$, but only 2 of these are independent because we impose 4~constraints: 
\begin{enumerate}
    \item The integral of the PDF over all $s$ is unity.
    \item The PDF is continuous at the transition point $s_{\rm g}$.
    \item The  PDF is differentiable at $s_{\rm g}$ (i.e., $dp/ds$ is continuous). 
    \item The total mass is conserved, so $\int_{-\infty}^{\infty} e^s p(s) = 1$.
\end{enumerate}

In fitting this functional form to the PDFs produced by our simulations, we leave the slope of the PL part ($\alpha_{\rm g}$) and the width of the LN part ($\sigma_{\rm s}$) as free parameters, and solve for the remaining parameters from the 4~constraints; details of the calculation can be found in \aref{sec:AppendixLNPL}. It is important to note that the well-known relation $s_0 = -\sigma_{\rm s}^2/2$ (see, e.g., \citealt{Federrath2008}) holds only for a pure LN PDF, and does not hold for a LN+PL form.

We also introduce a double power-law form of the density PDF, which we show below (\autoref{sec:Mach5case}) is a better fit to our simulations. This double power-law form, to which we shall refer as LN+2PL, is a straightforward extension of the LN+PL model, but now with two power-law sections,
\begin{equation}
\label{eq:defDPL}
 p(s) = \begin{cases}
 \frac{N}{\sqrt{2 \pi \sigma_{\rm s}^2}} \exp\left[-\frac{(s-s_0)^2}{2 \sigma_{\rm s}^2}\right] & s<\sg,  \\
N p_0 \exp\left(-s \alphag\right)  & \sg \leq s \leq \sd, \\
N p_1 \exp\left(-s \alphad\right)  & \sd \leq s \leq s_{\rm sink},
\end{cases}
\end{equation}
where $\alphad$ is the slope of the second PL tail, $\sd$ is the (log) density at which the first power law changes over to the second, $s_{\rm sink}$ is a maximum cut-off density and all other symbols have the same meanings as in the LN+PL case.\footnote{Note that, in contrast to the LN+PL form, for LN+2PL we include an explicit truncation of the PDF at $s = s_{\rm sink}$; this is necessary because, for many of the fits we will perform below, we obtain $\alpha_{\rm d} < 1$, in which case allowing the PDF to continue all the way to $s = \infty$ causes the normalisation to diverge. This is not necessary for the LN+PL models, because we can simply impose the constraint $\alpha_{\rm g} > 1$, and we find that in practice our best fitting values never approach this limit. However, this difference in how we handle the upper limit matters only for the normalisation, since in practice when we carry out fits to either the LN+PL or LN+2PL forms, we only fit the data on cells with $s<s_{\rm sink}$.} This model has nine free parameters, $N$, $\sigma_{\rm s}$, $s_0$, $s_{\rm g}$, $s_{\rm d}$, $p_0$, $p_1$, $\alpha_{\rm g}$, and $\alpha_{\rm d}$, out of which five are constrained by requiring the function
\begin{enumerate}
    \item be normalized, $\int_{-\infty}^{s_{\rm sink}} p(s) ds=1$, where $s_{\rm sink}$ is the sink particle creation threshold,
    \item conserve mass, $\int_{-\infty}^{s_{\rm sink}} e^s p(s)ds=1$,
    \item be continuous at $\sg$
    \item be continuous at $\sd$, and
    \item be differentiable at $\sg$.
\end{enumerate}
%(i) be normalized, $\int_{-\infty}^{s_{\rm sink}} p(s) ds=1$, where $s_{\rm sink}$ is the sink particle creation threshold, (ii) conserve mass, $\int_{-\infty}^{s_{\rm sink}} e^s p(s)ds=1$, (iii) be continuous at $\sg$ and $\sd$, and (iv) be differentiable at $\sg$. 
We choose to fit the four parameters $\sigma_{\rm s}$, $\alphag$, $\alphad$ and $\sd$, while determining the others from the constraints -- see \aref{sec:AppendixLNPLPL}.

The mechanics of fitting either the LN+PL or LN+2PL functions to a distribution of cell densities from a simulation require some care. One option would be to bin the simulation data in $s$, but this would produce an undesirable dependence of the results on the choice of binning. We therefore choose to fit using the cumulative distribution functions (CDFs) $P(s) = \int_{-\infty}^s p(s')\,ds'$ rather than the differential PDFs. In addition, to ensure that the PL tail(s) are fitted well, we use  
\begin{equation}{\label{eq:yPDF}}
    y(s) = \ln\left[\frac{P(s)}{1-P(s)}\right],
\end{equation}
rather than just $P(s)$ as our fitting function. When performing the fit, we omit cells with $s>s_{\rm sink}$, since in these cells the density distribution has been artificially altered by our sink particle method. Our code to carry out this fit is available publicly\footnote{\href{https://github.com/shivankhullar/PDF_Fit}{https://github.com/shivankhullar/PDF\_Fit}}.

\subsection{Relation between density PDF and star formation efficiency} \label{sec:epsff}

The star formation efficiency per free-fall time $\epsilon_{\rm ff}$ is a useful tool that has been measured and studied extensively on the scales of entire molecular clouds (\citealt{H10, L10, Krumholz14, Federrath2013b, SFK15, Onus18, Pokhrel2021}). 
\citet{Burkhart2018} demonstrates the importance of the form of the PDF and in particular its PL slope for the dimensionless rate of star formation ($\epsff$).

For an entire cloud, $\epsilon_{\rm ff} \equiv\mbox{SFR}\times t_{\rm ff}/M_{\rm gas}$ (\citealt{KM05}; \citetalias{FK12}), where SFR is the total star formation rate inside the cloud, $t_{\rm ff}$ is the cloud's free-fall time, and $M_{\rm gas}$ is the total gas mass; intuitively, $\epsilon_{\rm ff}$ is the ratio of the actual star formation rate to the rate that would be expected if the gas were in free-fall collapse. \citet{Khullar19} showed that one can obtain a useful diagnostic for simulations by extending this definition to consider only the material above some specified (log) density $s$,
\begin{equation}
\label{eq:epsff_vs_s_analytical}
    \epsilon_{\rm ff}(s) = \frac{{\rm SFR} \times t_{\rm ff}(s)}{M_{\rm gas}(s)} = 
    \frac{\mbox{SFR}\times t_{\rm ff,0}}{M_{\rm tot}}
    \frac{\int_s^\infty e^{-s/2} p(s) ds}{\int_s^\infty p(s) \,ds \times  \int_s^\infty e^s p(s) \,ds},
\end{equation}
where $t_{\rm ff}(s)$ and $M_{\rm gas}(s)$ are the mean free-fall time and total mass of gas with density $>s$, and $t_{\rm ff,0} = \sqrt{3\pi/32G\rho_0}$ and $M_{\rm tot}$ are the mean density free-fall time and total gas mass, respectively.

\begin{figure}
    \centering
    \includegraphics[width=\columnwidth]{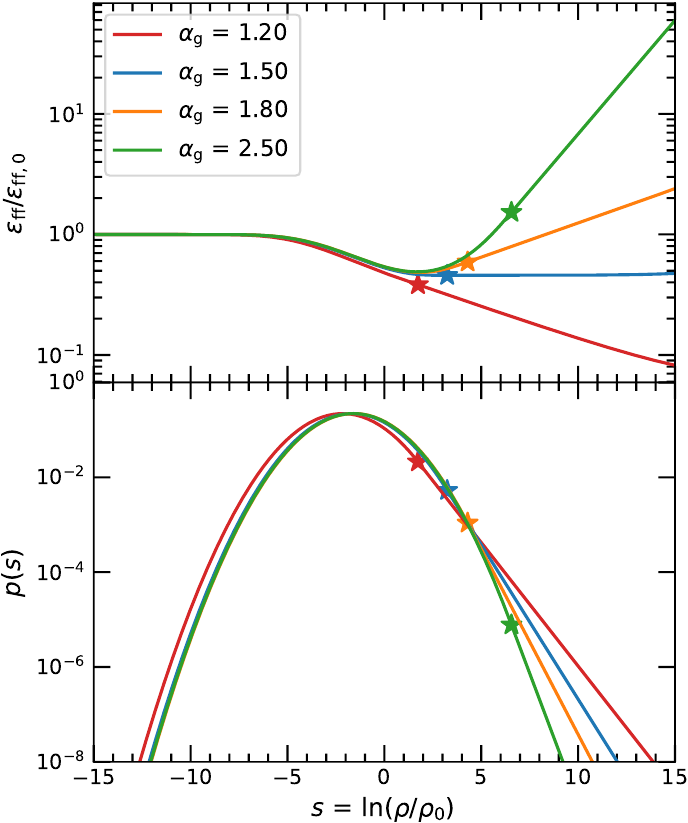}
    \caption{\textit{Bottom:} The density PDF described by \autoref{eq:SPL}, i.e., the combination of a lognormal and power law (LN+PL case). The different colors indicate PDFs having the same width of the LN part, $\sigma_{\rm s} = 1.8$, but different PL slopes $\alpha$. Star markers indicate the transition point, $\sg$, from the LN part to the PL part. \textit{Top:} The corresponding $\epsff(s)$ curve found using  \autoref{eq:epsff_vs_s_analytical}. While it may be difficult to distinguish the PL part from the tail of the LN part, the $\epsff(s)$ curves are very distinct and very sensitive to changes in the PL slope.} 
    \label{fig:PDFEff_analytical}
\end{figure}

In \autoref{fig:PDFEff_analytical}, we show sample density PDFs $p(s)$ and their corresponding $\epsilon_{\rm ff}(s)$ curves for LN+PL models with different PL slopes. We normalize the $\epsff$ curves by $\epsilon_{\rm ff,0} = \lim_{s\to -\infty}\epsff(s)$, which eliminates the constants SFR, $t_{\rm ff,0}$, and $M_{\rm tot}$. \autoref{fig:PDFEff_analytical} shows that the slope of the power-law tail of $\epsilon_{\rm ff}(s)$ is set by the density PDF slope $\alpha_{\rm g}$. For $\alpha_{\rm g} = 1.5$, we approach constant $\epsilon_{\rm ff}(s)$ as $s\to\infty$. The PL slope can be related to a density profile as $\rho(r)\propto r^{-3/\alphag}$ (\citealt{Kritsuk2011, FK13}).
For $\alphag=1.5$, $\rho(r) \propto r^{-2}$ which is typically seen during the early stages of gravitational collapse (\citealt{Larson1969, Penston1969a, Shu1977}).   

A PDF slope shallower than $1.5$ indicates slower than free-fall collapse, so there is an accumulation of mass at high densities. 
For the PL part, the slope of $\epsff(s)$ is given as 
\begin{equation}
    \frac{d\epsff(s)}{ds} = \frac{(3/2-\alphag)\alphag(\alphag-1)}{(\alphag+1/2)} e^{(\alphag+\frac{1}{2})s},
\end{equation}
which goes to zero when $\alphag=1.5$. 
%Note that for our density PDF models, $\alphag > 1$ is a constraint (see Appendix \autoref{sec:AppendixLNPL} for more details).

\section{Results}
\label{sec:Results}

Throughout this section, we will analyse all results as a function of star formation efficiency (SFE), in order to study the effects of cloud evolution. The SFE is defined as 
\begin{equation}
    {\rm SFE} (t) = \frac{M_* (t)}{M_*(t) + M_{\rm gas}(t)}, 
\end{equation}
where $M_*(t)$ and $M_{\rm gas}(t)$ are the mass in sink particles and gas, respectively, at any given time $t$. Note that in our simulations, the denominator $M_*(t)+M_{\rm gas}(t)=M$, the (constant) total cloud mass.

\subsection{Physical processes controlling the density PDF}
\label{sec:Mach5case}

Here we analyse the density structure, PDF and $\epsilon_{\rm ff}$ for simulation v1M5 ($\avir=1, \mathcal{M}=5$), to better understand the basic physical processes that control the shape of the density PDF. In \autoref{sec:PDFParamVariation} we analyse the dependence of the PDF on SFE and on the Mach number and virial parameter of the clouds.

\subsubsection{The density PDF $p(s)$ and its relation to the $\epsilon_{\rm ff}(s)$ curve}

\begin{figure}
    \centering
    \includegraphics[width=\columnwidth]{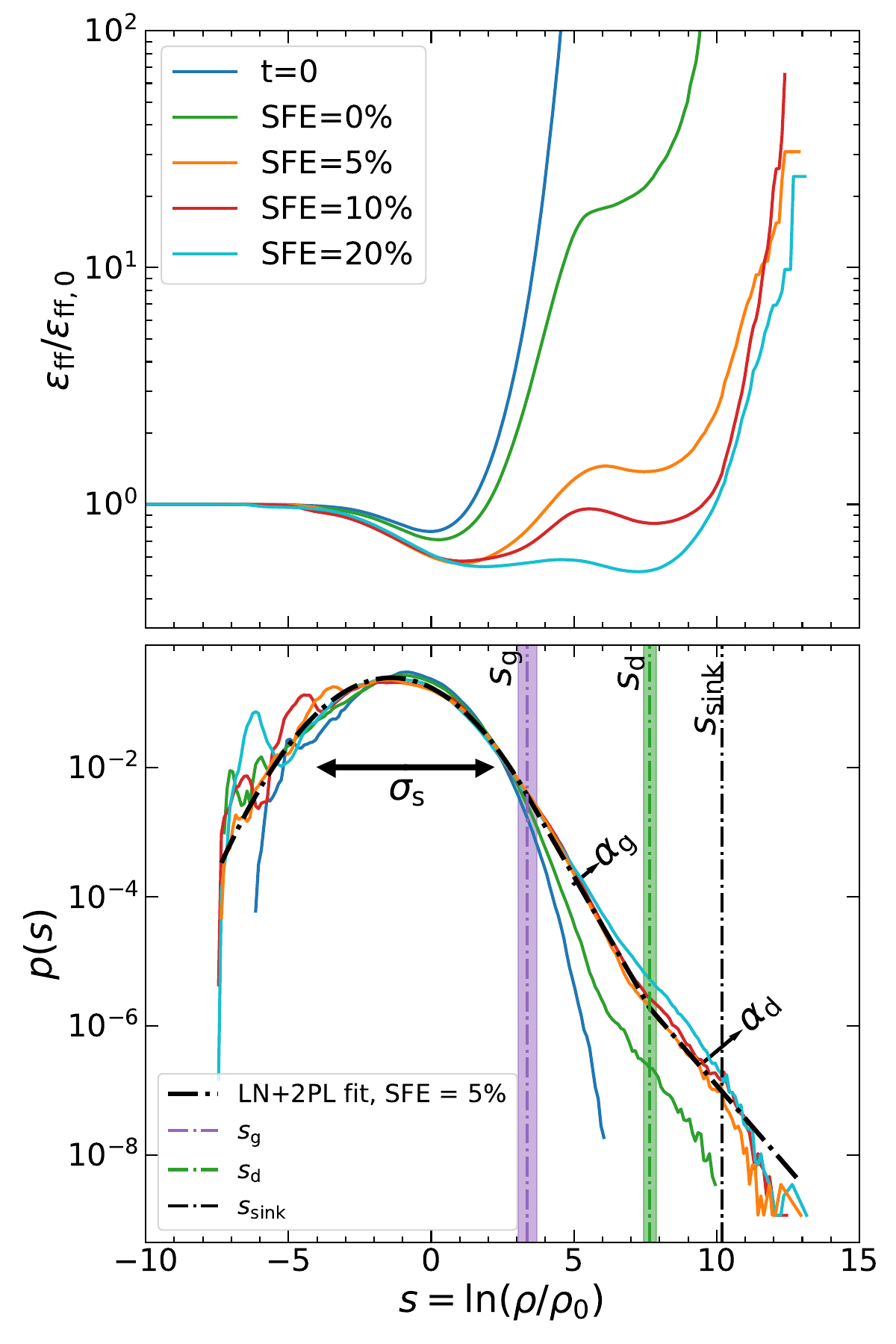}
    \caption{Dimensionless star formation efficiency $\epsilon_{\rm ff}(s)$ (\textit{top}) and density PDF $p(s)$ (\textit{bottom}) as a function of logarithmic density $s$ for the v1M5 simulation ($\avir=1, \mathcal{M}=5$) at different times during the evolution of the cloud. The different lines correspond to times $t=0$ when gravity is switched on (blue line), SFE = 0\%, just before the formation of the first sink particle (green line), SFE = 5\% (orange line), SFE = 10\% (red line) and SFE = 20\% (cyan line). A LN+2PL fit (\autoref{eq:defDPL}) to the PDF at SFE = 5\% is shown as the black dash-dotted line. The vertical dash-dotted lines in the bottom panel indicate the transition points between the LN part and the first PL, $\sg$ (purple), from the first PL to the second PL, $\sd$ (green), and at which sink particles form (black). The shaded regions correspond to the error bars in the fitted parameters.}
    \label{fig:Mach5fig}
\end{figure}

As discussed in \autoref{sec:epsff}, the density PDF $p(s)$ can be used to calculate the corresponding $\epsff(s)$ curve. 
In \autoref{fig:Mach5fig} we show the density PDF (bottom panel) and the $\epsff(s)$ curve (top panel) at different times during the evolution of the cloud in the v1M5 simulation. At $t=0$ when gravity is switched on, the shape of the PDF is close to a lognormal (LN) distribution (\citealt{Passot1998, Klessen2000, Kritsuk2007, Federrath2008, Federrath2010}), typical of supersonic turbulence. At SFE = $0$, just before the formation of the first sink particle, the width of the LN part of the PDF has broadened and the PDF has deviated enough from the LN to form a power-law (PL) tail; \citet{Jaupart2020} report this phenomenon as well. This tail has a slope close to $\alpha=1.5$, as is evident from the flat portion in the $\epsff(s)$ curve between $s\sim5$ and $s\sim7$. 

At $\mbox{SFE}\gtrsim 5\%$, the PDF is relatively steady and exhibits only minor variations in the slope of the PL tail(s). In this steady state, $\epsff(s)$ is fairly flat up to densities close to the sink particle threshold, where it artificially rises for numerical reasons, as mass is removed from the simulation grid and added to the sink particles. However, superimposed on this mostly flat behaviour, there is a distinct bump and dip structure, with a weak local maximum at $s\sim5$ and a minimum at $s\sim 7$. In terms of the density PDF, this corresponds to changes in the PL slope, from a steeper PL between $s\sim3-7$ to a shallower one at $s\gtrsim7$. 

\begin{figure}
    \centering
    \includegraphics[width=\columnwidth]{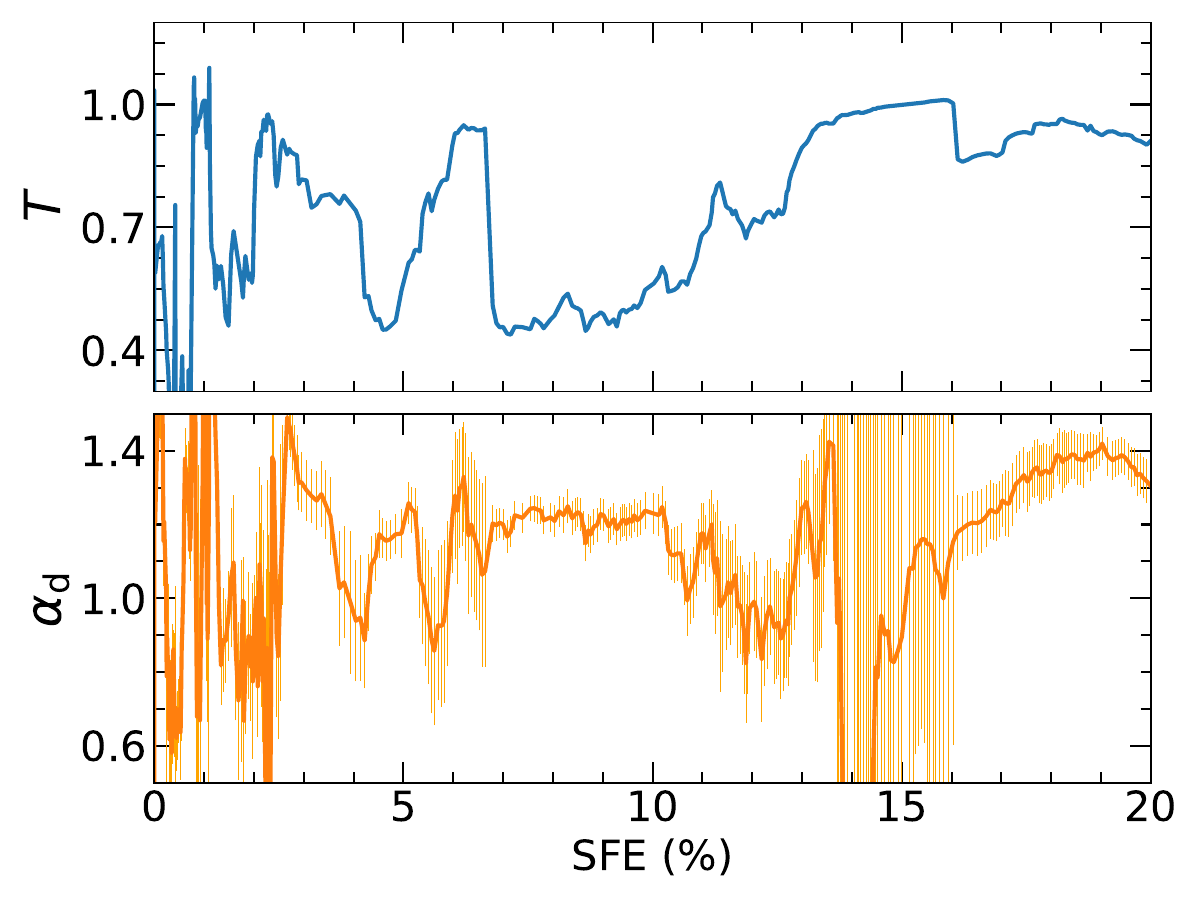}
    \caption{
    \textit{Top:} The quantity $T$, which is the ratio $\chi^2_{\rm red, LN+2PL}/\chi^2_{\rm red, LN+PL}$, as a function of SFE for the simulation v1M5. For large parts of the simulation, the LN+2PL model is a significantly better fit than the LN+PL model. \textit{Bottom:} The fitted values of the slope of the second PL tail, $\alphad$, with the vertical orange lines indicating the error bars on the estimated values. When $\chi^2_{\nu, {\rm 2PL}}/ \chi^2_{\nu, {\rm 1PL}}$ is close to 1, the LN+2PL density PDF parameters have large error bars or are highly variable.}
    \label{fig:modelcomparison}
\end{figure}

We can demonstrate quantitatively that the a LN+2PL model is a better fit to the simulation PDF than a LN+PL, at least once star formation begins, by comparing the reduced $\chi^2$ goodness-of-fit statistics for the two functional forms. We carry out LN+PL and LN+2PL fits to the density PDFs measured for every snapshot of simulation v1M5, and we plot the ratio of the reduced $\chi^2$ values, which we denote
\begin{equation}
T \equiv \frac{\chi^2_{\rm red,LN+2PL}}{\chi^2_{\rm red, LN+PL}}  
\end{equation}
as a function of SFE in the top panel of \autoref{fig:modelcomparison}. We find that $T<1$ for almost the entire simulation, indicating that the LN+2PL model is a better description of the simulation PDF. There are, however, exceptions: periods in the simulation (e.g., at $\mbox{SFE}\approx 13-16\%$) when $T$ comes close to or exceeds unity, indicating that a LN+PL fit is equivalent or preferred. At these times, we find that the error bars on the second PL slope, $\alphad$, become very large, as illustrated in the lower panel of \autoref{fig:modelcomparison}. Nonetheless, because the LN+2PL form is preferred at almost all times, we will adopt this functional form for the purposes of all our analysis in the remainder of this paper.

\subsubsection{Gas structures at the power-law transition points $\sg$ and $\sd$}

\begin{figure*}
    \centering
    \includegraphics[width=\textwidth]{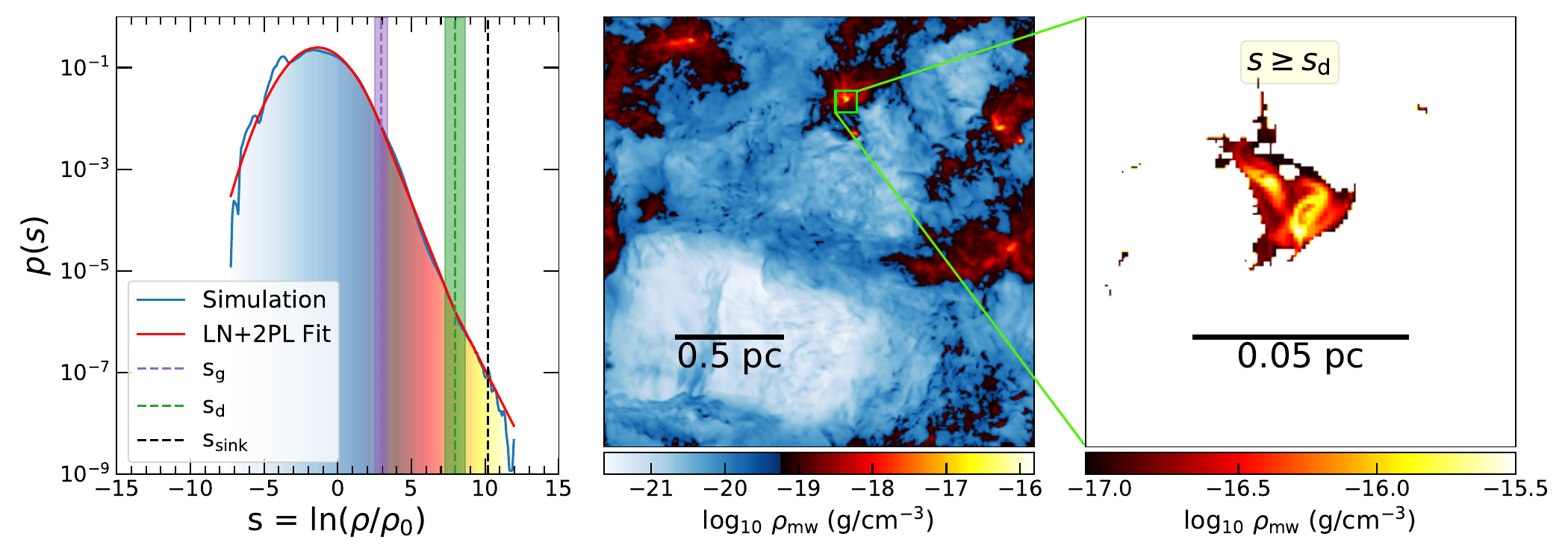}
    \caption{\textit{Left:} The density PDF for a snapshot from the simulation v1M5 ($\avir=1, \casemach5$) shown in blue, alongside our LN+2PL model fit (red). The vertical lines and corresponding shaded regions (errors in the fitted parameters) indicate the transition points from the LN part to the first PL, $\sg$ (purple), first PL to second PL, $\sd$ (green), and sink particle creation threshold density, $s_{\rm sink}$ (black). The change in the color-bar occurs at $s=\sg$ to indicate different regions in the gas flow. \textit{Centre:} The volume-weighted column density, $\rho_{\rm mw}$ (\autoref{eq:densweighteddens}), for the same simulation snapshot. The colorbar change occurs at $s=\sg$. \textit{Right:} A 0.1~pc box centred around the maximum density in the simulation box (green square in the middle panel). We only show gas that is denser the second transition point, $\sd$. The gas denser than $\sg$ consists of dense cores and filamentary-like structures, whereas gas denser than $\sd$ is almost always found in rotating discs.}
    \label{fig:discMach5only}
\end{figure*}

\begin{figure}
    \centering
    \includegraphics[width=\columnwidth]{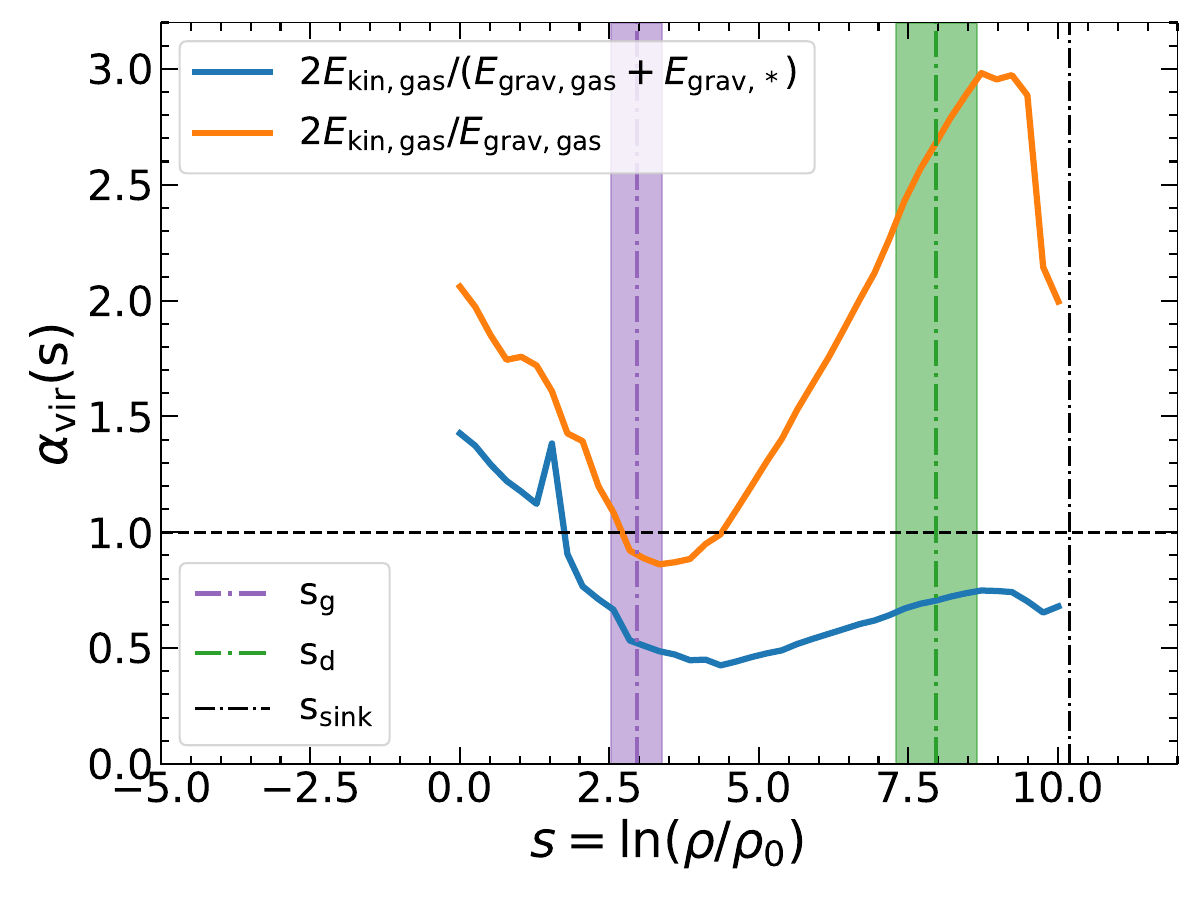}
    \caption{Virial parameter $\alpha_{\rm vir}(s)$ as a function of gas density $s$ for simulation v1M5 at SFE=6\%. The orange curve include the gravitational potential energy of the gas alone, whereas the blue curve includes the contributions of the stellar gravitational potential as well. As in \autoref{fig:discMach5only}, dash-dotted vertical lines and bands show the transition points and error bars in the best-fitting density PDF, $\sg$ (purple) and $\sd$ (green), and the sink particle threshold density $s_{\rm sink}$ (black). The horizontal dashed line indicates $\alphavir = 1$, which separates unbound gas ($\alphavir > 1)$ from gas that is bound and collapsing ($\alphavir<1)$. Gas denser than $\sg$ is bound and collapsing.}
    \label{fig:VirialParamFig}
\end{figure}

The LN+2PL model has two breakpoints: $\sg$, where the LN part transitions to the first PL (or, equivalently, the $\epsilon_{\rm ff}(s)$ curve begins to bend down), and $\sd$, where the first power law converts to the second, and the $\epsilon_{\rm ff}$ curve has its second inflection point. We next attempt to understand the nature of these transition points. We begin by looking at the column density morphology at different gas densities in \autoref{fig:discMach5only}, which shows the state of the v1M5 simulation at SFE = 6\%. In the left panel we show the density PDF and the corresponding LN+2PL model fit to the simulation data. The middle panel shows the mass-weighted mean density of the gas through the simulation volume, given by 
\begin{equation}
\label{eq:densweighteddens}
    \rho_{\rm mw} = \frac{\int \rho^2 \, dx}{\int \rho \,dx},
\end{equation}
where $x$ is the line-of-sight distance. We show the mass-weighted density projection because it picks out the dense gas structures along the line of sight better than a volume-weighted projection. The change in the colourbar from white-blue to black-red-yellow marks the transition point $\sg$. 

The right panel shows a zoomed-in 0.1~pc box centred around the maximum density (green box in the middle panel) and only shows gas denser than $\sd$.

From the figure, it is immediately obvious that the higher density threshold, $\sd$ is picking out structures found in rotationally-flattened discs. The connection between the second PL and rotationally-supported discs had previously been made by \cite{Kritsuk2011} and \cite{MurrayChangMurray2017}, and our analysis is consistent with theirs. The meaning of $\sg$ is somewhat less obvious from the morphology, but can be clarified by examining the physical state of the gas.

Specifically, we calculate the virial parameter $\alphavir$ (dropping the subscript to distinguish it from the virial ratio of the whole simulation box, $\avir$) as a function of gas (log over-)density, $s$. The virial parameter is  defined as the ratio of two terms in the virial theorem: twice the kinetic energy, divided by the absolute value of the potential energy: $\alphavir = 2E_{\rm kin}/|E_{\rm grav}|$. Both $E_{\rm kin}$ and $E_{\rm grav}$  depend on the structure and distribution of the gas. To find $\alphavir (s)$, we  extract connected regions above a density $s$, which we decompose into clumps using  {\texttt{yt}}'s {\texttt{Clump Finder}}\footnote{\href{https://yt-project.org/doc/analyzing/analysis_modules/clump_finding.html}{https://yt-project.org/doc/analyzing/analysis\_modules/clump\_finding.html}} \citep{Turk2011}.  We compute the kinetic  energy for each clump in its centre-of-mass frame, and we compute the gravitational potential both for the gas alone, and for the combined gravity of the gas and the stars.\footnote{For the gravitational energy, we use the full  potential due to all gas and stars within the simulation, rather than just the self-gravity of the region in the clump.  This has the effect of suppressing $\alphavir$ somewhat for clumps deep in an external potential well (such as overdensities within an accretion disc). However, this is a minor effect, which does not change our qualitative conclusions.} 

We plot $\alpha_{\rm vir}(s)$ in \autoref{fig:VirialParamFig}. The figure shows that $\alpha_{\rm vir} > 1$ for densities comparable to $\rho_0$ ($s\approx 0$), regardless of whether we include the stellar contribution to the potential. As the density increases, $\alphavir$ decreases, reaching a minimum $\alphavir<1$ at $s\approx s_{\rm g}$, indicating that this gas is bound and collapsing; at this transition, stellar gravity contributes roughly half the total potential energy. As we go to even higher densities, $\alphavir$ rises until it reaches $s\approx \sd$. The rise is sharp if we neglect stellar gravity, but only very shallow if we include it. It is  therefore clear that stellar gravity largely dictates the behaviour of the gas density PDF for $s\geq\sg$. At densities higher than $\sg$, $\alphavir$ increases slightly (while still being $<1$ throughout), primarily due to increased turbulent and rotational support as we approach the disc density $\sd$. It reaches a maximum at $s\approx \sd$, before flattening out.

We are now in a position to interpret the features of the PDF: $\sg$ marks the transition between bound and unbound gas, and at this transition gas and stellar gravity are roughly equally important. For $s > \sg$, stellar gravity becomes increasingly dominant, such that the gas is not bound to itself, but is bound to stars. At $s \approx \sd$, rotational support becomes increasingly important, such that gas at $s \gtrsim \sd$ is found largely in rotationally-supported discs.

\subsection{Variation of the density PDF parameters}
\label{sec:PDFParamVariation}
Having described our density PDF model and the physical processes controlling it, we now look at how the parameters describing the PDF vary over time and as a function of Mach number and (global) virial parameter. 

\subsubsection{Dependence of the PDF on SFE}

\begin{figure*}
    \centering
    \includegraphics[width=\textwidth]{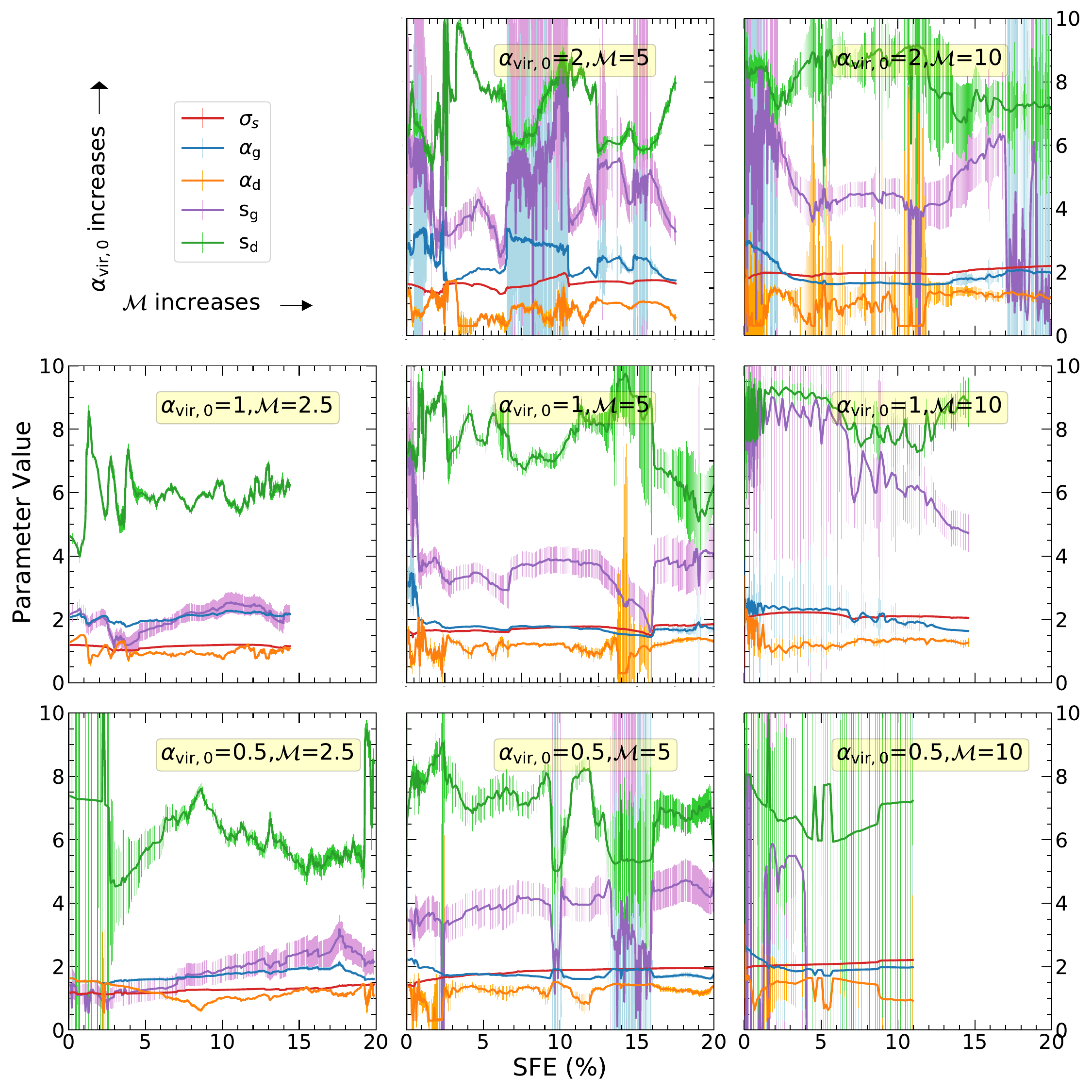}
    \caption{The evolution of the various density PDF parameters: (a) the width of the LN part of the PDF, $\sigma_{\rm s}$ (in red), (b) the two PL slopes, $\alphag$ and $\alphad$ (blue and orange respectively), (c) the two transition points, $\sg$ and $\sd$ (purple and green respectively), as a function of Star Formation Efficiency (SFE) in different panels for different simulation parameters. The heavy central line shows the best-fitting value, and the vertical lines around them show the $1\sigma$ uncertainty. Various trends are discussed in detail in \autoref{sec:PDFParamVariation}. The parameters $\sigma_{\rm s}$ and $\sg$ tend to increase with increasing SFE, whereas the PL slopes $\alphag$ and $\alphad$ remain fairly constant.}
    \label{fig:all_params}
\end{figure*}

We fit our LN+2PL model described in \autoref{sec:Methodology} to every snapshot in each of our simulations. In \autoref{fig:all_params}, we show how the six important (four fitted, two derived) PDF parameters vary with SFE for each of the simulations. The plot is arranged so that each row has a constant $\avir$ and varying $\mach$, whereas each column has constant $\mach$ and varying $\avir$.

Starting with the width of the LN part of the PDF, $\sigma_{\rm s}$ (red lines), we note that $\sigma_{\rm s}$ appears to be nearly constant throughout the star formation process in most of the simulations. The only secular trend visible is that $\sigma_{\rm s}$ increases slightly with SFE in the simulations with $\alpha_{\rm vir,0} = 0.5$. For these simulations, the gravitational collapse tends to widen the distribution of the LN part of the PDF, as the turbulence becomes increasingly gravity-driven and thus compressive\footnote{If we consider the time between $t=0$ and $\mbox{SFE}=0\%$, all the simulations show an increase in the width of the LN part of the PDF in this duration.}. This is consistent with the predictions of \citet{Jaupart2020}; we discuss comparisons with this model, and extensions to it, further below. 

%At $t=0$, before gravity is switched on, the width of the LN PDF agrees very well with the relation $\sigma_{\rm s}^2 = ln(1+b^2\mach^2)$ (cite cite). 

The PL slopes, $\alphag$ (blue lines) and $\alphad$ (orange lines), remain fairly constant with SFE for all simulations, independent of $\mach$ or $\avir$. As already shown in \autoref{fig:modelcomparison}, there exist certain windows of time (or SFE) for which a LN+PL model fits the data almost as well as our fiducial LN+2PL, and these are reflected in the intermittently large error bars on the estimates of $\alphag$ or $\alphad$. Nonetheless, it is clear that there are no secular trends in $\alphag$ or $\alphad$ with SFE.

The transition points between the LN and the first PL, $\sg$ (purple lines), and the first PL and the second PL, $\sd$ (green lines), exhibit more interesting behaviour. In the $\casemach2.5$ and $\casemach5$ simulations, considering the range of SFEs where $\sg$ is well defined (i.e., ignoring times when $\sg$ has large error bars because the PDF is ambiguous between LN+PL and LN+2PL), we find that $\sg$ tends to increase with SFE. This is especially true for the $\avir=0.5$ cases, since gravity plays a more important role for them. This trend is not surprising given that $\sg = s_0 + \alphag \sigma_{\rm s}^2$ (see \autoref{eq:sgSPL}), and that $\sigma_{\rm s}$ increases slightly with SFE, whereas $\alphag$ and $s_0$ do not show a systematic increase or decrease with SFE. Our estimation of the break point between the first and second PL, $\sd$, reveals that for the $\casemach2.5$ and $\casemach5$ cases, there are relatively large fluctuations, which make it hard to identify any trends across the range of SFEs probed here. For the $\casemach10$ simulations, our resolution is not sufficient to resolve the discs in detail, especially for the $\avir=0.5$ and $\avir=2$ cases (see \aref{sec:AppendixdiscSizeResCrit} for more details on the resolution criterion). Consequently, the uncertainties on $\sg$ and $\sd$ are very large throughout the simulations, which precludes us from analysing these cases further.

\begin{figure*}
    \centering
    \includegraphics[width=\textwidth]{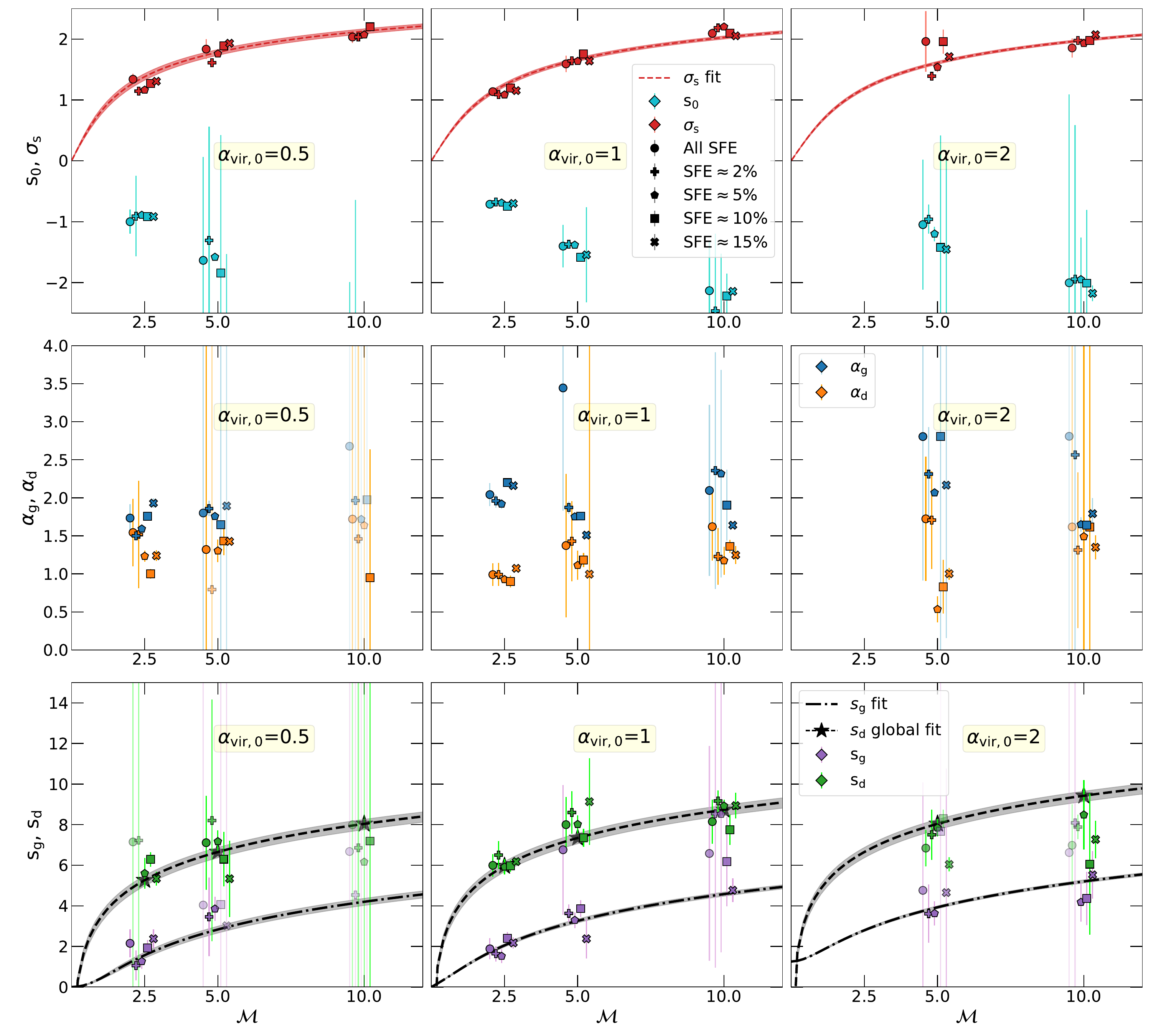}
    \caption{The variation of the PDF parameters on the simulation parameter $\mach$, keeping $\avir$ constant. (\textit{Top row:}) $\sigma_{\rm s}$ (red) and $s_0$ (cyan), (\textit{Middle row:}) $\alphag$ (blue) and $\alphad$ (orange), (\textit{Bottom row:}) $\sg$ (purple) and $\sd$ (green), with the Mach number $\mach$, for 3 different virial parameters ($\avir=0.5$ (left column), $\avir=1$ (centre column), $\avir=2$ (right column). The different shapes of the points indicate the average over an SFE range; plus shapes represent the average between SFE=1\% and 3\%, pentagons between SFE=4\% and 6\%, squares between SFE=9\% and 11\%, crosses between SFE=14\% and 16\% (or <16\% for simulations stopped before then), and lastly, circles represent the average over the entire SFE range. 
    The red dashed line and the dash-dotted line in the top and bottom row respectively represent our fit to \autoref{eq:Effective_b_sigma} and \autoref{eq:Effective_b_sGJC20} for each of the 3 different cases $\avir=0.5,1,2$ (see \autoref{sec:Effective_driving}). The dashed line in the bottom row shows the global fit to our model for the disc density (see \autoref{sec:disc-calculation}, \autoref{eq:disc-thresh}). The shaded regions indicate the 2$\sigma$ error bars for each of our fits. Various trends are discussed in \autoref{sec:DependenceMachVir}.
    }
    \label{fig:Params_mach}
\end{figure*}

\begin{figure*}
    \centering
    \includegraphics[width=\textwidth]{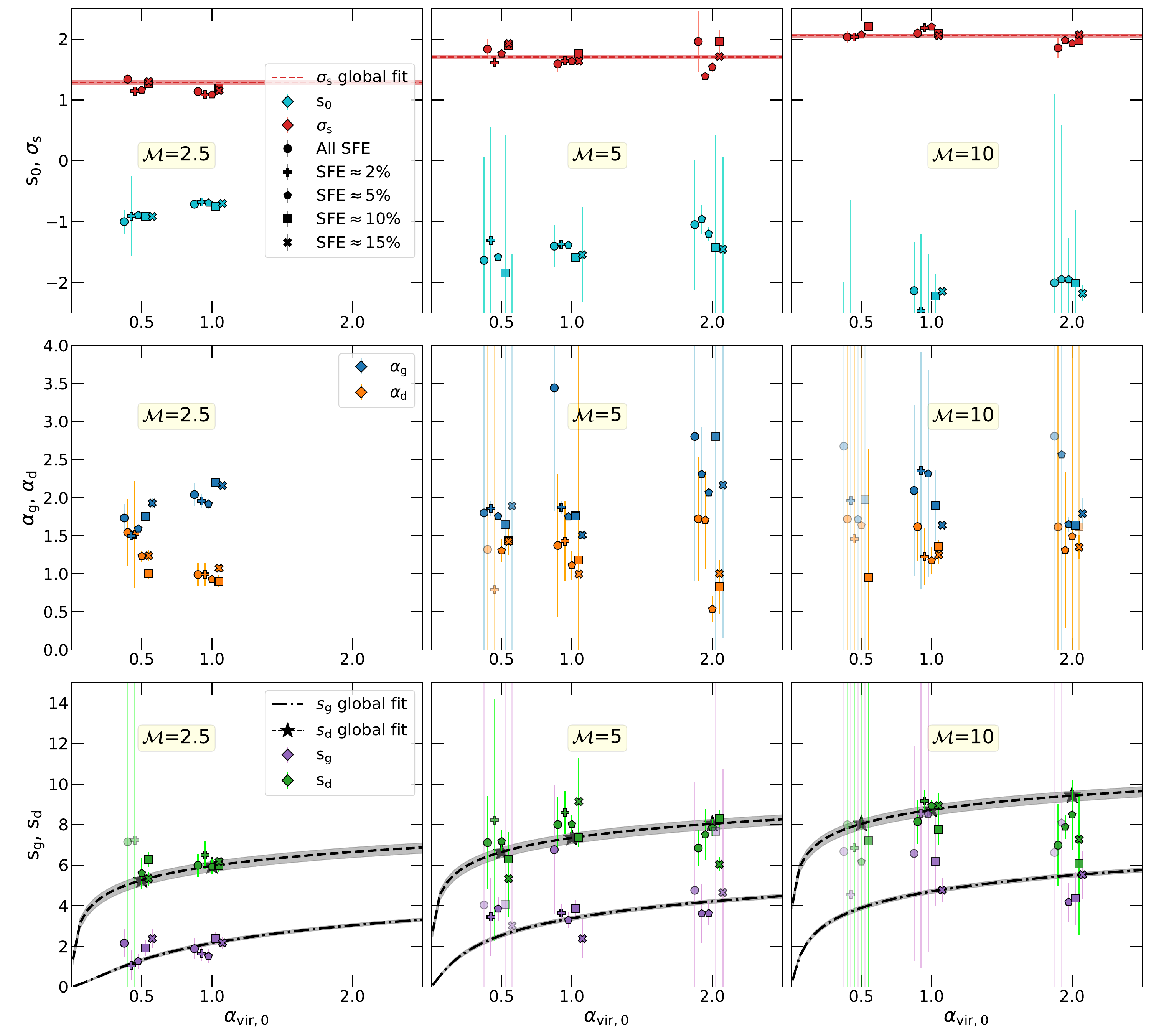}
    \caption{The variation of the PDF parameters on the simulation parameter $\avir$, keeping $\mach$ constant. (\textit{Top row:}) $\sigma_{\rm s}$ (red) and $s_0$ (cyan), (\textit{Middle row:}) $\alphag$ (blue) and $\alphad$ (orange), (\textit{Bottom row:}) $\sg$ (purple) and $\sd$ (green), with the simulation virial parameter $\avir$, for 3 different Mach numbers ($\casemach2.5$ (left column), $\casemach5$ (centre column), $\casemach10$ (right column). The different shapes of the points indicate the average over an SFE range; plus shapes represent the average between SFE=1\% and 3\%, pentagons between SFE=4\% and 6\%, squares between SFE=9\% and 11\%, crosses between SFE=14\% and 16\% (or <16\% for simulations stopped before then), and lastly, circles represent the average over the entire SFE range. 
    The red dashed line and the dash-dotted line in the top and bottom row respectively represent our global fit to \autoref{eq:Effective_b_sigma} and \autoref{eq:Effective_b_sGJC20} for all the simulations (see \autoref{sec:Effective_driving}). The dashed line in the bottom row shows the global fit to our model for the disc density (see \autoref{sec:disc-calculation}, \autoref{eq:disc-thresh}). The shaded regions indicate the 2$\sigma$ error bars for each of our fits. Various trends are discussed in \autoref{sec:DependenceMachVir}.}
    \label{fig:Params_vir}
\end{figure*}

\subsubsection{Dependence of the PDF on the Mach number and virial parameter}
\label{sec:DependenceMachVir}

We now look at the dependence of the fitted PDF parameters on $\mach$ and $\avir$, since ultimately, any efforts to systematically study the density PDF should enable us to understand how its shape varies as a function of quantities measurable through observations. For this purpose, for each simulation we average the density PDFs for all snapshots over three intervals of SFE: 1\% -- 3\%, 9\% -- 11\%, and 14\% -- 16\% (or to the highest SFE attained, for simulations that do not reach 16\%). We then fit LN+2PL functional forms to these averaged PDFs, extract the best-fitting parameters, and plot the results in \autoref{fig:Params_mach} and \autoref{fig:Params_vir}; these two figures show the same underlying data, but the former shows variation with $\mach$ at fixed $\avir$, while the latter shows variation with $\avir$ at fixed $\mach$.

First consider the dependence on $\mach$. The top row of \autoref{fig:Params_mach} shows that the width and the peak of the LN part of the PDF, $\sigma_{\rm s}$ and $s_0$, clearly depend on $\mach$. This is to be expected given that, in the absence of self-gravity, the former obeys the well-known relationship  \citep{Federrath2008, Federrath2010, Price2011, Konstandin2012, Molina2012, FederrathBanerjee2015, Nolan2015, Kainulainen2017}
\begin{equation}
\label{eq:LNwidth}
\sigma_{\rm s}^2=\ln(1+b^2\mach^2).
\end{equation}
While this relationship predicts the correct qualitative trend of $\sigma_{\rm s}$ with $\mach$, quantitatively \autoref{eq:LNwidth} systematically under-predicts the measured value of $\sigma_{\rm s}$. To take one example, run V0.5M2.5 (top left panels in \autoref{fig:Params_mach} and \autoref{fig:Params_vir}) has $\mach = 2.5$, and for our natural driving mix, we expect $b\approx 0.4$; plugging in, the predicted value of $\sigma_{\rm s} \approx 0.8$, whereas the values that we obtain by fitting the observed PDF are $\approx 1.1-1.2$, roughly 30\% larger. We will return to the source of this undershoot below.

The peak of the PDF, $s_0$, also varies with $\mach$, which is again expected since conservation of mass requires $s_0 = -\sigma_{\rm s}^2/2$ if the PDF is a pure LN. For our LN+2PL model, $s_0$ does not have a straightforward dependence on $\sigma_{\rm s}$ (see \autoref{sec:AppendixLNPLPL}), but is still strongly anti-correlated with it, and consequently with $\mach$. 

The middle row of \autoref{fig:Params_mach} shows the two PL slopes, $\alphag$ and $\alphad$, as a function of $\mach$. 
%The slopes dictate how gas behaves in different stages of star formation. 
From the figure it is difficult to identify any clear trends with $\mach$, especially since the uncertainties on $\alphag$ and $\alphad$ are relatively large. This is because, while the LN+2PL model is a better fit to the data most of the time, there are times when a single PL is a better description of the density PDF, and during these intervals the values of the slopes become highly uncertain.

%For the single PL model, the PL slope tends to get shallower with increasing SFE (also noted by CITE). This apparent change in the single PL slope, is likely a combination of two PLs developing with differing slopes, as we find for most of the simulations studied here. In the theory presented by \cite{Jaupart2020}, the authors suggest that there is an initial PL slope of $\alpha\approx2$, but then a second PL tail with $\alpha=1.5$ emerges when regions are in free-fall collapse. While this is true for the initial stages of collapse, this work shows that over longer timescales (steady-state condition), the two PL slopes have approximately similar values of about 1.5 and the PL tail with a slope shallower than 1.5 is a result of gas spending many dynamical times rotating in an accretion disc around the star before being accreted. The presence of magnetic fields or feedback process which control the physics of these discs will also likely dictate the values of the PDF parameters pertaining to the second PL tail, $\alphad$ and $\sd$.

In the bottom row of \autoref{fig:Params_mach} we show the dependence of $\sg$ and $\sd$ on $\mach$. We see a clear trend for $\sg$ to increase with $\mach$, also seen in \citet{Burkhart2019}, which is expected given that $\sg \propto \sigma_{\rm s}^2$ (see \aref{sec:AppendixLNPLPL}). The increased width of the LN for higher $\mach$ pushes $\sg$ to higher $s$. Since $\alphag$ does not show systematic trends, the trend in $\sg$ with $\mach$ is determined solely by how $s_0$ and $\sigma_{\rm s}$ vary. We return to this variation and the physical reasons for it in \autoref{sec:Effective_driving}. %\cite{Jaupart2020} give an expression for estimating $\sg$ as 
%\begin{equation}
%\label{eq:sGJC20}
%    |e^{\sg}-1| = (b\mach)^2 \times \avir \times \left| \frac{\sg+\frac{1}{2}\sigma_{\rm s}^2}{\sigma_{\rm s}^2} \right|,
%\end{equation}
%which predicts an increase in $\sg$ with an increase in $\mach$.
We also find that $\sd$ increases with $\mach$, with the exception of the $\casemach10$ case for which the separation of the two scales $\sg$ and $\sd$ is limited by numerical resolution constraints (see \aref{sec:AppendixdiscSizeResCrit}).

We now consider the dependence of the fitted PDF parameters on the simulation virial parameter, $\avir$, in \autoref{fig:Params_vir}. Once again, the top panel shows how $\sigma_{\rm s}$ and $s_0$ vary with $\avir$, the middle panel shows the variation of $\alphag$ and $\alphad$, and the bottom panel shows the variation of $\sg$ and $\sd$ with $\avir$. There is a weak anti-correlation of $\sigma_{\rm s}$ with $\avir$ such that increasing $\avir$ decreases $\sigma_{\rm s}$. The reason for this is easy to understand: in the low-$\avir$ cases, the dynamics are determined mostly by self-gravity, and this results in an increase in the width of the LN part of the density PDF because it makes the turbulence more compressive (as discussed above in the context of \autoref{fig:all_params} above). The dependence of $s_0$ on $\avir$ can be understood similarly: $s_0$ increases with increasing $\avir$ since $s_0$ and $\sigma_{\rm s}$ are anti-correlated. The virial parameter $\avir$ has no significant effect on the PL slopes, $\alphag$ and $\alphad$.  We do not expect any such dependence, since they are set by how the process of collapse proceeds in self-gravitating gas, and this is fairly independent from the varying strength of turbulence relative to self-gravity on larger scales. However, we caution that our ability to draw any strong conclusions is once again limited by the large error bars on our fitted PL slopes. 

Lastly, the transition points, $\sg$ and $\sd$, do not show any strong dependence on $\avir$, except a weak increase with $\avir$ for the well-resolved cases (i.e., except for $\mach=10$). \citet{Burkhart2019} expect a weak dependence of $\sg$ on $\avir$. We will seek to explain these trends in the next section.

\section{Physical models for the PDF parameters}
\label{sec:Physical_models}

%One of the useful traits of the density PDF (and the N-PDF) is the dependence of the PDF parameters on the physical cloud parameters such as $\mach$ and $\avir$. Several efforts have been made 
Several authors have proposed models
to relate cloud properties to density PDF parameters. For instance, the slopes of the PL tails, $\alphag$ (or $\alphad$) can be related to the density profile of the collapsing gas cloud as $\rho(r) \propto r^{-3/\alphag}$ (e.g., \citealt{FK13, Kritsuk2011}). In this section we shall seek to explain the trends observed in \autoref{sec:DependenceMachVir} for the width of the LN part and the values of the two transition points.

\subsection{Lognormal width and first powerlaw transition}
\label{sec:Effective_driving}

As noted earlier, in the presence of gravity, the width of the LN part of the density PDF, $\sigma_{\rm s}$, increases, as does the log density $\sg$ at which the LN gives way to powerlaw behaviour. As discussed above, the former trend is qualitatively consistent with the relationship between $\sigma_{\rm s}$ and $\mach$ that has long been known for non-self-gravitating turbulence (\autoref{eq:LNwidth}). The latter trend is also qualitatively consistent with the recent model proposed by \citet{Jaupart2020}, who predict a relationship
\begin{equation}
\label{eq:sGJC20}
    |e^{\sg}-1| = (b\mach)^2 \times \avir \times \left| \frac{\sg+\frac{1}{2}\sigma_{\rm s}^2}{\sigma_{\rm s}^2} \right|.
\end{equation}
where in \citeauthor{Jaupart2020}'s model $b\mach$ and $\sigma_{\rm s}$ are functions of simulation time $t$ to be obtained by fitting the low-density part of the PDF; such fitting is needed because, as several authors have shown, self-gravity increases the effective driving parameter, so we cannot assume that $b$ late in the simulation is the same as it was before the onset of collapse \citep[e.g.,][]{ Jaupart2020, Kortgen2020, Menon2020, Menon2021}. While the resulting values of $\sg$ would be consistent with our data, this approach does not constitute a full \textit{a priori} prediction of $\sg$, since it requires as input values of $b\mach$ and $\sigma_{\rm s}$ that are measured from the PDF, rather than predicted. Since we have found that the PDF is approximately stationary once the star formation efficiency reaches a few percent, this motivates us to instead attempt to predict the transition point $\sg$ solely in terms of the simulation parameters $\mach$ and $\avir$, without relying on explicit fits to the time-dependent PDF. To differentiate between the initial value of $b$ linked to the type of turbulent forcing used in our simulations and its evolved value due to gravity, we re-write \autoref{eq:LNwidth} as
\begin{equation}
    \label{eq:Effective_b_sigma}
    \sigma_{\rm s}^2 = \ln(1+b_{\rm eff}^2\mach^2)
\end{equation}
where $b_{\rm eff}$ is an effective driving parameter in the presence of gravity, which is to be determined by fitting a simple function of $\mach$ and $\avir$; the resulting value of $\sg$ is then
\begin{equation}
\label{eq:Effective_b_sGJC20}
    |e^{\sg}-1| = (b_{\rm eff}\mach)^2 \times \avir \times \left| \frac{\sg+\frac{1}{2}\sigma_{\rm s}(b_{\rm eff})^2}{\sigma_{\rm s}(b_{\rm eff})^2} \right|,
\end{equation}
where $b_{\rm eff}$ is the predicted effective driving parameter, and $\sigma_{\rm s}(b_{\rm eff})$ is the value given by \autoref{eq:Effective_b_sigma}.

To investigate whether this 
%single-parameter modification is sufficient to explain the discrepancies between the models and the data
simple model is sufficient, we perform a non-linear least-squares fit for $b_{\rm eff}$ using the values of $\sigma_{\rm s}$ and $\sg$ predicted by \autoref{eq:Effective_b_sigma} and \autoref{eq:Effective_b_sGJC20}, respectively, and those obtained from the LN+2PL fits to our simulations; for the purposes of this fit, we weight errors in $\sigma_{\rm s}$ and $\sg$ equally, so the best fit is the value of $b_{\rm eff}$ that minimises the sum of the squared difference between the predicted and observed values of $\sigma_{\rm s}$ and $\sg$. We carry out this fit in two ways; first, we fit different values of $b_{\rm eff}$ for the three cases $\avir=0.5,1,2$, and second we fit all simulations together; we refer to this latter case as a global fit. We show the results of the first fitting method in \autoref{fig:Params_mach}, and the second in \autoref{fig:Params_vir}. (Recall that the underlying data are the same in the two figures; they are simply organised differently.) We report the best-fitting values of $b_{\rm eff}$ for both cases in \autoref{tab:b_param_fit}.

Examining the figures, it is clear that either the global or the case-by-case fits yield predicted values of $\sigma_{\rm s}$ and $\sg$ that are in reasonably good agreement with the measurements, and capture both the correct mean values and the dependence on $\mach$; the model-predicted dependence on $\alphavir$ is also consistent with the simulation measurements, although in this case our limited dynamic range in $\alphavir$ and the noisiness of the simulation results means that no dependence on $\alphavir$ at all would probably be almost an equally-good model.

%Nonetheless, this experiment demonstrates that
%our simple procedure of 
%modifying the value of $b$ seems to accomplish our goal of relating the transition point to the simulation parameters without . 
%most of the problems in the models. 
\autoref{tab:b_param_fit} shows that the values of $b_{\rm eff}$ %require\textbf{d} for this purpose
predicted by either the case-by-case or global versions our linear model are significantly greater than the $b=0.4$ expected for a natural mix of compressive and solenoidal modes, and are substantially closer to the $b=1$ expected for purely compressive turbulence. In the case where we allow different values of $b_{\rm eff}$ for different $\alphavir$, the effective driving parameter shows a clear anti-correlation with the virial parameter of the simulation. As the relative strength of gravity increases ($\avir$ decreases), the effective driving parameter increases, consistent with the idea that, in these simulations, gravity is providing a substantial contribution to the turbulent velocity, and that it drives predominantly compressive modes (as suggested by \citealt{Jaupart2020}). This experiment demonstrates that our simple linear model for $b_{\rm eff}$ accomplishes our goal of predicting the transition point solely from the simulation parameters, without relying on an explicit fit to the LN part of the density PDF.
We acknowledge, however, that this is just one possible %explanation for the discrepancies between the models and the data, one that we have adopted largely on the basis of simplicity and the principle of making minimal modifications to existing models. 
approach.
Other modifications to \autoref{eq:LNwidth} are certainly plausible, and with our limited and noisy sampling of parameter space, we are not in a position to rule out such approaches.
%There could exist other possible parametrisations, including a breakdown of the usual Mach-variance relation (\autoref{eq:LNwidth}) altogether.

\begin{table}
    \centering
    \begin{tabular}{c|c} \hline
         \textbf{Case} & $b_{\rm eff}$ \\ \hline \hline
          $\avir=0.5$ & 0.95$\pm$0.04 \\
          $\avir=1.0$ & 0.77$\pm$0.02 \\
          $\avir=2.0$ & 0.70$\pm$0.01 \\ \hline
          Global fit & 0.82$\pm$0.02 \\ \hline
    \end{tabular}
    \caption{Best-fitting effective turbulence driving parameter $b_{\rm eff}$ from \autoref{eq:Effective_b_sigma} and \autoref{eq:Effective_b_sGJC20}, obtained using the data in \autoref{fig:Params_mach}. In the top three rows, we show fits obtained by separately fitting the simulations with $\alphavir=0.5,1$, and 2, as indicated; these 3 cases correspond to 3 columns in \autoref{fig:Params_mach}. The bottom row, Global fit, shows the value of $b_{\rm eff}$ obtained by fitting for all the simulations simultaneously.}
    \label{tab:b_param_fit}
\end{table}

%. However, in the presence of gravity, the width of the LN part of the PDF increases and we shall interpret this as an increase in the driving parameter $b$. We therefore modify \autoref{eq:LNwidth} and \autoref{eq:sGJC20} by replacing $b$ with this effective driving parameter $b_{\rm eff}$. We then use these modified equations simultaneously to fit for $b_{\rm eff}$ from our simulation data for each of the 3 different cases $\avir=0.5,1,2$. We show the results of these fits in \autoref{fig:Params_mach} and summarize our  

\subsection{Disc transition scale}
\label{sec:disc-calculation}

Having determined the scalings of $\sigma_{\rm s}$ and $\sg$ with cloud properties, we next seek to do the same for the characteristic density of disc formation, $\sd$. For this purpose we adapt an argument from \citet{BurkertBodenheimer2000}, in which motions at the onset of collapse are assumed to sample the background line width-size relation.  We estimate the scale of a collapsing region to be the sonic scale of the turbulence cascade.  Approximate angular momentum conservation then yields a characteristic disc scale, and the importance of gravitational instability determines the outer disc density.  Similar estimates appear in previous works by \citet{MatznerLevin05} and \citet{KratterMatzner06}, and this calculation is a minor extension of the one presented by \citet{Burkhart2019}.  

We begin  with the line-width size relation 
\begin{equation}
    \sigma_{V}(r) = \sigma_c \left( \frac{r}{R_c} \right)^\beta,
\end{equation}
where $R_c$ is the driving scale for turbulent motions, $\sigma_c = \sigma_V(R_c) = \mach c_s$.  Across the supersonic turbulent cascade, $\beta\approx1/2$, and this suffices for our purposes despite the fact that the exponent tends toward Kolmogorov value of ($\beta=1/3$) around the sonic transition \citep{Fedderath2021}.  Here   $r_s$ is the sonic scale, at which $\sigma_{V}(r_s) = c_s$; therefore 
\begin{equation}
\label{eq:sonicscale}
    r_s = \frac{R_c}{\mach^{1/\beta}}.
\end{equation}

To collapse, a region on this scale must achieve a gas density $\rho_s$  above the critical density required for collapse: 
\begin{equation}
    r_s^2 = \frac{\phi_1 c_s^2}{G \rho_s}
\end{equation}
where $\phi_1$ is a factor of order unity. This implies a critical sonic mass given by $m_s = \phi_2 \rho_s r_s^3$, where we account for geometry and complicated dynamics with the factor $\phi_2$. %In the spirit of  \citet{MatznerLevin05} and \citet{KratterMatzner06}, w
We define the disc radius as $r_d = \phi_d r_s$; here $\phi_d$ is significantly smaller than unity, and reflects the typical turbulent angular momentum of a region on the brink of collapse. 

The disc density can be written as $\rho_d = \phi_3 \Sigma_d \Omega/c_s = \phi_3 \Omega^2/(\pi G Q)$, where $\Sigma_d$ is the surface density, $\Omega=\phi_4 G m_s/r_d^3$ is the orbital speed and $Q=c_s \Omega/(\pi G \Sigma_d)$ is the \citet{Toomre64} stability parameter. Using our definition for the virial parameter (eq. \ref{eq:vir-analytic}), and adopting  $R_c = L/2$, 
%$M_c=\rho_0 L^3$, we can simplify to obtain, 
we obtain 
\begin{equation}
\label{eq:disc-thresh}
    e^{\sd} = \phi_w  \frac{ \avir \mach^{\frac{2}{\beta} -2}}{Q}
\end{equation}
where % $Q\approx1$, $\phi_w$ is defined as 
$\phi_w$ is a collection of the other coefficients, defined as 
  %$ 24 \phi_v/(5\pi)$, $\phi_v = \phi_u / \phi_d^3$ and $\phi_u = \phi_1\phi_2\phi_3\phi_4$.
$ 24 \phi_1\phi_2\phi_3\phi_4/(5\pi \phi_d^3)$.  We expect $Q\approx 1-2$ so that gravitational instability stimulates the disc angular momentum transport.

%\footnote{\textbf{Note that some aspects of this derivation are similar to \citealt{Burkhart2019}, however, their derivation is for the transition from the LN to the PL and we have included more parameters to account for the disc transition.}}

While every one of these dimensionless quantities may depend on the dimensionless  physical and numerical parameters of our simulations (and presumably does), we hypothesize that the dependence is weak so long as the disc scale is reasonably well resolved.   In this spirit, we perform a one parameter fit of our model to the estimates of $\sd$ from the simulations. For this, we fix $\beta=1/2$ and allow $\phi_w$ to be a free parameter.
%, whereas for the two parameter fit, we allow the exponent of $\mach$ (q=2/$\beta$-2) in eq \ref{eq:disc-thresh} to vary freely as well.
We find that a global fit to all our simulations results in $\phi_w = 62.0 \pm 7.6$. This would mean that the best fit value for the product $\phi_1\phi_2\phi_3\phi_4\approx0.13$, which is reasonable.
%We find that the values of $\phi_u$ are reasonable since the product $\phi_1\phi_2\phi_3\phi_4$ could produce factors of $\approx 0.13$. 
We show the predictions using this best-fitting model in \autoref{fig:Params_mach} and \autoref{fig:Params_vir}. While there is a significant amount of scatter (at least in part because $\sd$ is not always well-determined in our simulations), we see that the model captures both the absolute value of $\sd$ and its dependence on $\alphavir$ and $\mach$ reasonably well.
\section{Conclusions}
\label{sec:Conclusions}

Self-gravitating turbulence is a common description of the flow in star-forming molecular clouds, and a common paradigm for theories of star formation. However, we still lack a complete theory for the density distributions produced in such media. In this work we characterise the density distribution in self-gravitating, non-magnetised, turbulent media by performing numerical simulations with a range of Mach numbers and virial parameters. Using these simulations, we show that:
\begin{enumerate}
\item The volumetric density PDF can be divided into three regimes: turbulence dominated, gravity dominated, disc/rotation dominated. These regimes are characterised by different functional forms: the PDF transitions from being a lognormal (LN) at low densities (turbulence-dominated regime) to a first powerlaw (PL; gravity-dominated) and a second (typically shallower) powerlaw (disc regime) at higher densities. 
Our LN+2PL model includes elements of the LN+PL model by \citet{Burkhart2018} and \citet{Burkhart2019}, in particular, the importance of the sonic scale in setting a PL transition. However, contrary to their results, we find for the most part that the PDF does not evolve with star formation efficiency (SFE), and instead remains nearly steady once the SFE reaches a few percent. 
%we find that the intermediate PL is important and exists over a broad range of densities, and for this reason the PDF is not completely specified by the Mach number and the sonic scale.

\item While the break in the power-law (PL) slope can be difficult to see directly in the PDF, statistical tests confirm that a fit containing two power laws is almost always preferred over one with a single power law, and the presence of a second power law becomes obvious when we examine the star formation efficiency $\epsilon_{\rm ff}(s)$ as a function of log density $s$ \citep{Khullar19}. In this diagnostic, the break between the gravity-dominated and disc-dominated regimes manifests as a characteristic oscillatory pattern whereby $\epsilon_{\rm ff}(s)$ reaches a local maximum in the gravity-dominated regime and then a local minimum at the transition from this regime to the disc regime. This oscillatory pattern is present in all of our simulations at almost all times once star formation begins. 

We note that the analytical model of \citet{Jaupart2020} includes a lognormal with two power law tails: a first transient PL with a steep slope ($\sim2$) covering the density range over which gravity strongly affects turbulent acceleration, and a second PL with an index of index of  $\sim1.5$ representing gravity-dominated dynamics. \citet{Jaupart2020} predict that the second PL tail moves steadily to lower densities, ultimately replacing most of the first power law after about a free-fall time.  
%\skcut{Our two PL regions resemble these predictions (albeit with consistently shallower second PL, $\alphad\lesssim1.5$) -- with the important exception that our second PL  represents rotationally supported material.}
However, our two PLs differ from theirs: our first PL covers roughly the same density range over which they fit two PLs, and has a slope $\alpha_{\rm g}$ between 1.5 and 2, i.e., intermediate between the slopes of their two proposed PLs. By contrast our second PL appears at much higher density and corresponds to rotationally-supported material. It is only visible in simulations that at least marginally resolve disc (Appendix~\ref{sec:AppendixdiscSizeResCrit}), which the simulations used by \citealt{Jaupart2020} do not. %\skadd{These two PLs differ, however, with our two PLs. Our first PL with $\alphag=1.5$ corresponds to the second regime described by \citealt{Jaupart2020} while our second PL corresponds to rotationally supported material.} \skcomment{I don't think this last statement is true because our $\alphag$ is not equal to 1.5} 
    \item The gravity dominated part of the PDF is characterised by gas with local virial parameter $\alphavir<1$ (in agreement with observational results of \citealt{Chen2018, Chen2019}). This is a result of stellar gravity being more important as we go from $\sg$, the log density at which gravity becomes dominant and where gas self-gravity and stellar gravity are about equally important, to $\sd$, the log density at which rotational support becomes dominant, where only stellar gravity is relevant. At densities above $\sd$ rotational support kicks in, leaving $\alphavir\sim1$, and the morphology becomes disc-like.
    \item The width of the LN part of the PDF increases with Mach number, and scale at which the LN gives way to a gravity-dominated PL increases with both Mach number and virial parameter. The rates of increase are consistent with prior theoretical predictions, provided we make one substantial modification to these models: self-gravitating turbulence is significantly more compressive than turbulence driven with a natural mix of modes, which has the effect of increasing both the LN width and the transition density. Models predicting these two parameters yield reasonably accurate results only if we set the turbulence-driving parameter $b$ that appears in them to a value close to that expected for purely compressive turbulence, rather than one that reflects the properties of the turbulence that is intially present at the onset of star formation. 
    \item We show that the disc formation scale, and its variation with Mach number and virial parameter, can be understood as arising from the characteristic amount of angular momentum at the sonic scale of the turbulence.
\end{enumerate}

We therefore conclude that in a turbulent self-gravitating medium, the nature of gas flows are captured in the shape of the density PDF and that a LN+2PL model can be used to describe this density PDF well. With the help of this model, we can determine the effects of the molecular cloud properties on the density PDF, a necessary step toward formulating a predictive theory of star formation. However, we caution that we have not yet included the effects of stellar feedback or magnetic fields, which would likely modify the functional form of the PDF. We leave these steps to future work, and caution against applying the conclusions of our study directly to observed molecular clouds until we are able to carry them out. 
%\textbf{The inclusion of magnetic fields and stellar feedback would likely impact the shape and evolution of the density PDF and alter the dependence of the PDF parameters on the SFE in addition to changing how SFE increases with time (\citealt{Collins2012, Sadavoy2014, F15, Schneider2015, Lin2016, Mocz2017, Grudic2018}).}

\section*{Acknowledgements}

We would like to thank the anonymous referee for their comments. SK would like to thank Norman Murray and Piyush Sharda for valuable discussions. We further thank Etienne Jaupart and Gilles Chabrier for their comments on the manuscript. The research of SK and CDM is supported by an NSERC Discovery Grant. MRK acknowledges support from the Australian Research Council through its Future Fellowship and Discovery Projects funding schemes, awards FT180100375 and DP190101258. CF acknowledges funding provided by the Australian Research Council (Future Fellowship FT180100495), and the Australia-Germany Joint Research Cooperation Scheme (UA-DAAD). Computations were performed on the Niagara supercomputer at the SciNet HPC Consortium. SciNet is funded by: the Canada Foundation for Innovation; the Government of Ontario; Ontario Research Fund - Research Excellence; and the University of Toronto. We further acknowledge high-performance computing resources provided by the Australian National Computational Infrastructure (grant~ek9) in the framework of the National Computational Merit Allocation Scheme and the ANU Merit Allocation Scheme, and by the Leibniz Rechenzentrum and the Gauss Centre for Supercomputing (grant~pr32lo). The simulation software FLASH was in part developed by the DOE-supported Flash Center for Computational Science at the University of Chicago.
\textit{Greenhouse gas emissions}: We estimate 140kg of CO$_2$ equivalent from simulations presented in this work, based on $\sim$ 4470 kWh of Toronto, ON, electricity.
\section*{Data availability}

The simulation data underlying this article will be shared on reasonable request to the corresponding author.

%%%%%%%%%%%%%%%%%%%%%%%%%%%%%%%%%%%%%%%%%%%%%%%%%%



\bibliographystyle{mnras}
\bibliography{References.bib} % if your bibtex file is called example.bib

%\begin{thebibliography}{99}
%\end{thebibliography}

%%%%%%%%%%%%%%%%%%%%%%%%%%%%%%%%%%%%%%%%%%%%%%%%%%

%%%%%%%%%%%%%%%%% APPENDICES %%%%%%%%%%%%%%%%%%%%%

\appendix

\section{The LN+PL and LN+2PL models}
\label{sec:AppendixFitting}


In this section, we describe our LN+PL and LN+2PL models in more detail, including all the constraints and the relations between the parameters, and how we use these constraints when we carry out fits.

\subsection{LN+PL}
\label{sec:AppendixLNPL}

Starting with the definition of our function, $p(s)$ (see \autoref{eq:SPL}), we first require that the integral over all $s$ be normalized to unity. This constrains the normalization parameter $N$ to be
\begin{equation}
\label{eq:NSPL}
     N = \frac{1}{\frac{p_0 e^{-\alphag \sg}}{\alphag}+\frac{1}{2} \text{erfc}\left(\frac{s_0-\sg}{\sqrt{2} \sigma_{\rm s}}\right)}.
\end{equation}

Next, we apply the constraint for differentiability at the transition point, $\sg$. The derivatives on either side of $\sg$ are
\begin{equation}
\label{eq:DPL}
 p'(s) = \begin{cases}
   -\frac{(s-s_0) e^{-\frac{(s-s_0)^2}{2 \sigma_{\rm s}^2}}}{\sqrt{2 \pi } \sigma_{\rm s}^3 \left(\frac{p_0 e^{-\alphag \sg}}{\alphag}+\frac{1}{2} \text{erfc}\left(\frac{s_0-\sg}{\sqrt{2} \sigma}\right)\right)} & s-\sg<0 \\
 -\frac{\alphag p_0 e^{-\alphag s}}{\frac{p_0 e^{-\alphag \sg}}{\alphag}+\frac{1}{2} \text{erfc}\left(\frac{s_0-\sg}{\sqrt{2} \sigma}\right)} & s-\sg\geq 0.
\end{cases}
\end{equation}
These two expressions are equal if the parameter $p_0$ obeys
\begin{equation}
\label{eq:p0SPL}
    p_0 = -\frac{(s_0-\sg) e^{\alphag \sg-\frac{(\sg-s_0)^2}{2 \sigma_{\rm s}^2}}}{\sqrt{2 \pi } \alphag \sigma_{\rm s}^3}.
\end{equation}

Applying the continuity constraint and substituting for $p_0$, we get
\begin{equation}
\label{eq:sgSPL}
     \sg = s_0 + \alphag\sigma_{\rm s}^2,
\end{equation}
which is similar to the expression derived by \citet{BB17}, except that we cannot use the relation $s_0=-1/2 \sigma_{\rm s}^2$ that holds for pure LN PDFs, since we impose a constraint on mass conservation in our model. This final constraint reads $\int_{-\infty}^\infty e^s p(s) ds=1$. Thus we can solve for $s_0$, under the assumption that $\sigma_{\rm s}>0$ and $\alphag>1$, as 
\begin{equation}
\label{eq:s0SPL}
    s_0 = \log \left(\frac{A}{B}\right)
    %+2 i \pi  c_1,c_1\in \mathbb{Z}, 
\end{equation}
where
\begin{eqnarray}
A & = &(\alphag-1) \left[\sqrt{\pi } \alphag \sigma_{\rm s} e^{\frac{\alphag^2 \sigma_{\rm s}^2}{2}} \left(\text{erf}\left(\frac{\alphag \sigma_{\rm s}}{\sqrt{2}}\right)+1\right)+\sqrt{2}\right], \\
B & = & \sqrt{2} \alphag e^{\alphag \sigma_{\rm s}^2}
\nonumber \\
\lefteqn{
{} + \sqrt{\pi } (\alphag-1) \alphag \sigma_{\rm s} e^{\frac{1}{2} \left[\alphag^2+1\right) \sigma_{\rm s}^2} \left(\text{erf}\left(\frac{(\alphag-1) \sigma_{\rm s}}{\sqrt{2}}\right)+1\right].
}
%     B & = &\sqrt{\pi } (\alphag-1) \alphag \sigma_{\rm s} e^{\frac{1}{2} \left(\alphag^2+1\right) \sigma_{\rm s}^2} \left(\text{erf}\left(\frac{(\alphag-1) \sigma_{\rm s}}{\sqrt{2}}\right)+1\right) \\
%     & & {} +\sqrt{2} \alphag e^{\alphag \sigma_{\rm s}^2}.
\end{eqnarray}
%Taking $c_1=0$, we can find $s_0$.
When fitting to LN+PL distributions, our procedure is to leave $\alpha_{\rm g}$ and $\sigma_{\rm s}$ as free parameters. We then derive $s_0$ from \autoref{eq:s0SPL}, $\sg$ from \autoref{eq:sgSPL}, $p_0$ from \autoref{eq:p0SPL}, and finally $N$ from \autoref{eq:NSPL}, which completes specification of the model.

\subsection{LN+2PL}
\label{sec:AppendixLNPLPL}

Starting with the definition of our function $p(s)$, defined in \autoref{eq:defDPL}, we impose the normalization constraint first. This gives, 
\begin{eqnarray}
\label{eq:NormLN2PL}
N & = &\left[ \frac{p_0}{\alphag} (e^{-\alphag \sg} - e^{-\alphag \sd}) + \frac{p_1}{\alphad} (e^{-\alphad \sd} - e^{-\alphad s_{\rm sink}})
\right.
\nonumber \\
& & \left.
{} 
+ \frac{1}{2} {\rm erfc} \left( \frac{s_0-\sg}{\sqrt{2} \sigma_{\rm s}} \right)  \right]^{-1}
%\begin{array}{rr}
%N & = \left[ \frac{p_0}{\alphag} (e^{-\alphag \sg} - e^{-\alphag \sd}) + \frac{p_1}{\alphad} (e^{-\alphad \sd} - e^{-\alphad s_{\rm sink}}) \right\\
%& + \left \frac{1}{2} {\rm Erfc} \left( \frac{s_0-\sg}{\sqrt{2} \sigma_{\rm s}} \right)  \right]^{-1}
%\end{array}
\end{eqnarray}
Next, we apply the differentiability constraint at the transition points $\sg$, which gives us
\begin{equation}
\label{eq:p0LN2PL}
    p_0 = \frac{\sg - s_0}{\sqrt{2\pi} \sigma_{\rm s}^3 \alphag} {\rm exp}\left( \alphag \sg - \frac{(\sg-s_0)^2}{2\sigma_{\rm s}^2} \right).
\end{equation}
As for the single PL case, we next apply the continuity constraint at $\sg$, which leads once again to
\begin{equation}
\label{eq:sgLN2PL}
    \sg = s_0 + \alphag \sigma_{\rm s}^2.
\end{equation}
We cannot require differentiability at the second transition point $\sd$ as that would imply $\alphag=\alphad$. However, we do require continuity at $\sd$, which gives us
\begin{equation}
\label{eq:p1LN2PL}
    p_1 =  p_0 e^{-(\alphag - \alphad)\sd}.
\end{equation}
Finally, we apply the mass conservation constraint. Evaluating the integral over the various parts of the PDF, this can be expressed as
\begin{align}
\label{eq:s02PL}
    \frac{N}{2}(J+K+L) = 1
\end{align}
where
\begin{eqnarray}
    J & = & \frac{2 p_0}{\alphag-1} [e^{(\sg - \alphag \sg)} - e^{(\sd - \alphag \sd)}], \\
    K & = &\frac{2 p_1}{\alphad-1} [e^{(\sd - \alphad \sd)} - e^{(s_{\rm sink} - \alphad s_{\rm sink})}], \\
    L & = & {\rm exp}\left( s_0 + \frac{\sigma_{\rm s}^2}{2} \right) \times {\rm erfc} \left( \frac{s_0 + \sigma_{\rm s}^2 - \sg}{\sqrt{2} \sigma_{\rm s}} \right).
\end{eqnarray}
When fitting LN+2PL functions, we leave $\sigma_{\rm s}$, $\alpha_{\rm g}$, $\alpha_{\rm d}$, and $\sd$ as our four free parameters. 
%\mrknote{Please confirm that my understanding of your numerical procedure is correct.} 
Since the equations above constitute a transcendental system for the remaining parameters, we must solve them numerically. Fortunately, we can reduce the problem to a one-dimensional root-finding exercise as follows: for any set of the four fitting parameters and a trial value of $s_0$, we can use \autoref{eq:sgLN2PL} to obtain $\sg$, \autoref{eq:p0LN2PL} to obtain $p_0$, \autoref{eq:p1LN2PL} to obtain $p_1$, and then \autoref{eq:NormLN2PL} to solve for $N$. In general the value of $N$ obtained thereby will not satisfy the mass constraint given by \autoref{eq:s02PL}.
We therefore iteratively adjust $s_0$ until this condition is satisfied, at which point we will have found the correct, converged values for $s_0$, $\sg$, $p_0$, $p_1$, and $N$.

%%%%%%%%%%%%%%%%%%%%%%%%%%%%%%%%%%%%%%%%%%%%%%%%%%

\section{Disc resolution criterion}
\label{sec:AppendixdiscSizeResCrit}
\begin{figure*}
    \centering
    \includegraphics[width=\textwidth]{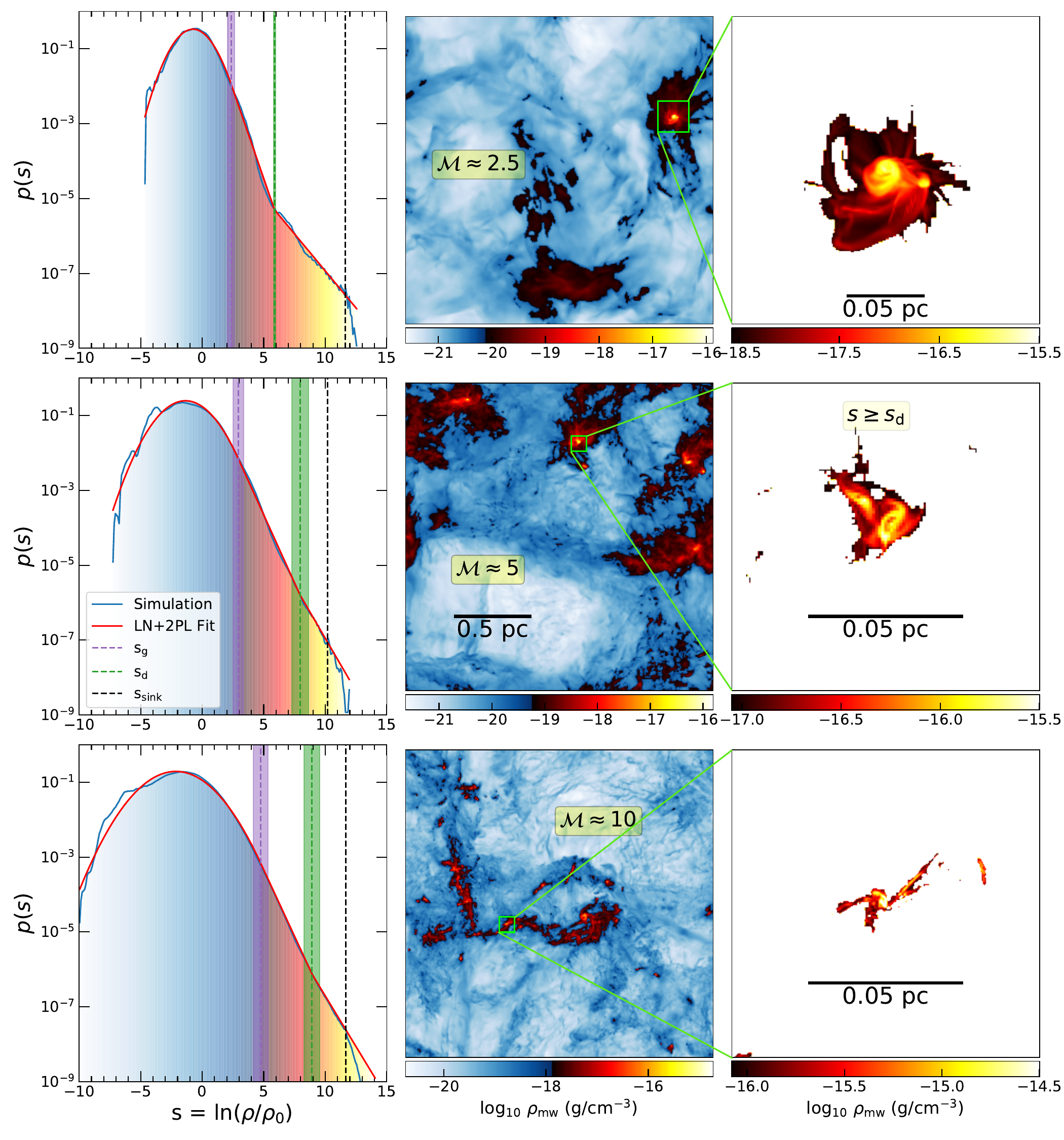}
    \caption{Each row represents simulations with 3 different Mach numbers, $\casemach2.5$ (\textit{top}), $\casemach5$ (\textit{middle}), $\casemach10$ (\textit{bottom}), but the same virial parameter, $\avir=1$. \textit{Left column:} The density PDF for a single snapshot shown in blue, alongside our LN+2PL model fit (red). The vertical lines and corresponding shaded regions (errors in the fitted parameters) indicate the transition points from the LN part to the first PL, $\sg$ (purple), first PL to second PL, $\sd$ (green), and sink particle creation threshold density, $s_{\rm sink}$ (black). The change in the color-bar occurs at $s=\sg$ to indicate different regions in the gas flow. \textit{Centre column:} The volume-weighted column density, $\rho_{\rm mw}$ (\autoref{eq:densweighteddens}), for the same simulation snapshots. The colorbar change occurs at $s=\sg$. \textit{Right column:} A 0.2~pc box ($\casemach2.5$) and 0.1~pc boxes ($\casemach5, \casemach10$) centred around the maximum density in the simulation boxes (green squares in the middle panel). We only show gas that is denser than the second transition point, $\sd$. The gas denser than $\sg$ consists of dense cores and filamentary-like structures, whereas gas denser than $\sd$ is almost always found in rotating discs. The disc sizes decrease as the $\mach$ number increases.}
    \label{fig:discMach}
\end{figure*}

\begin{figure}
    \centering
    \includegraphics[width=\columnwidth]{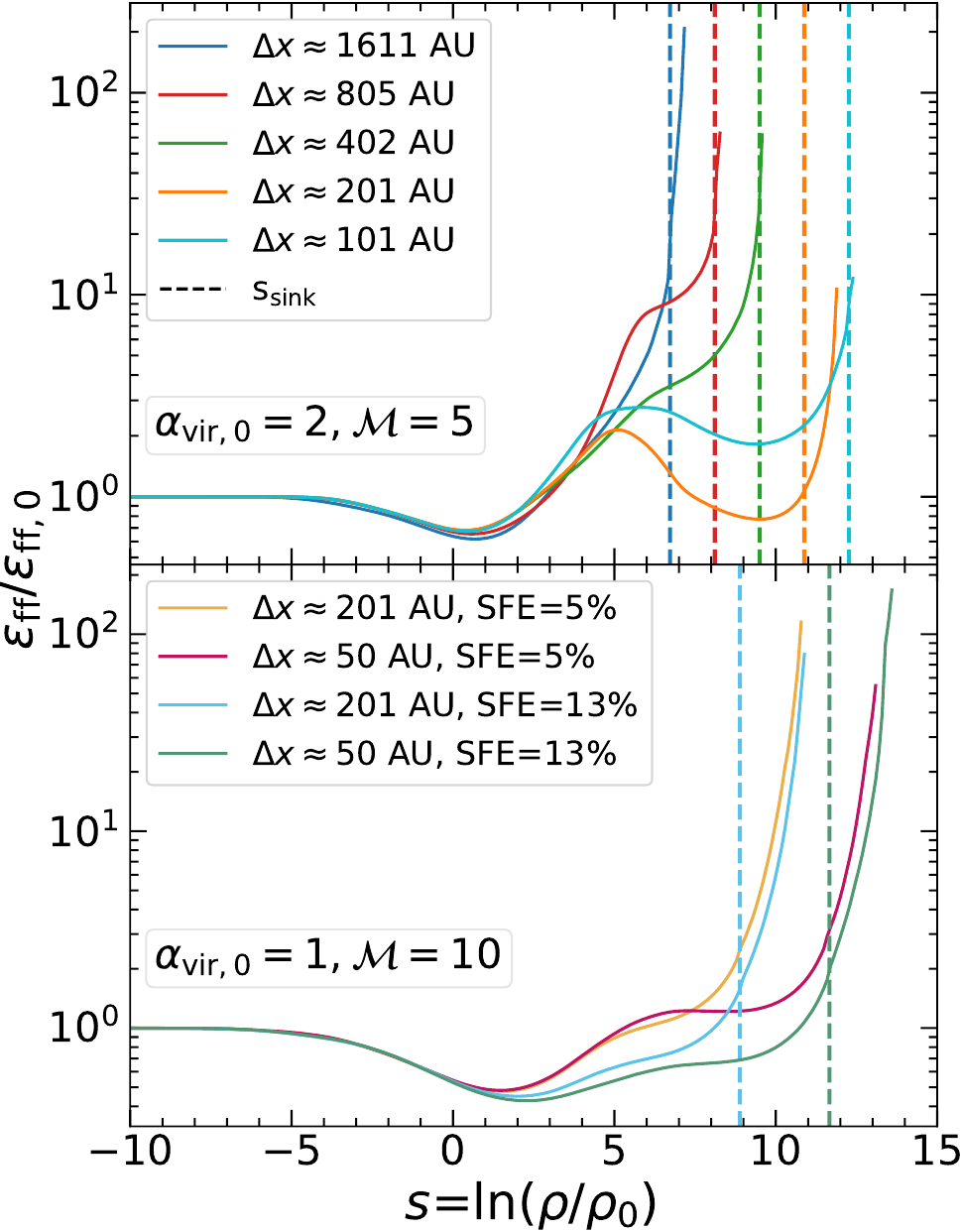}
    \caption{\textit{Top:} $\epsff(s)$ curves for the simulations with $\avir=2, \casemach5$ (case v2M5), but differing effective resolutions: 1611 AU (blue), 805 AU (red), 402 AU (green), 201 AU (orange), 101 AU (cyan). The curves are shown for an SFE=5\%, but are time-averaged over the range 4-6\%. The vertical dashed lines indicate the sink particle density threshold for the corresponding simulations. The slope of the $\epsff(s)$ curve depends on the PL tail slope in the density PDF and therefore can be used as a proxy for estimating the resolution required to resolve the second PL tail and disc-like structures. \textit{Bottom:} $\epsff(s)$ curves at different SFE for the simulations with $\avir=1, \casemach10$ (v1M10) but differing effective resolutions: 201 AU (orange and cyan), and 50 AU (red and teal). Once again, SFE=5\% represents a time-average over the SFE range 4-6\%, and SFE=13\% is a time-average over the range 12-14\%. The $\Delta x\approx50$ AU simulation curves start to show indications of the rise and fall structure that is indicative of a second PL tail, but do not quite reach the resolution that would be required to capture it.}
    \label{fig:ResdiscSize}
\end{figure}

In this appendix we investigate the resolution required for a simulation to capture the formation of discs and the transition from the first to second PL. \autoref{fig:discMach} is similar to \autoref{fig:discMach5only}, and is constructed following the same procedure; however, in \autoref{fig:discMach} each of the three rows shows a simulation with a different Mach number (top: $\casemach$ 2.5, middle: $\casemach$ 5, bottom: $\casemach$ 10) but the same virial parameter, $\avir=1$. The right-most column of \autoref{fig:discMach} shows the gas denser than the second transition point $\sd$, which lies almost entirely in discs. There is a clear dependence of the disc size on the Mach number, which is not unexpected, since in \autoref{eq:sonicscale} we show that disc sizes are a fraction of the sonic scale, $r_{\rm d} \propto \mach^{-2}$. This implies that a disc in a $\casemach5$ simulation is about 4 times larger than in a $\casemach10$ simulation, completely consistent with the ratio of disc sizes visible in \autoref{fig:discMach}. 

We next look at the resolution required to resolve the discs and as a result the second PL tail. As shown in \autoref{sec:epsff}, the $\epsff(s)$ curve is very sensitive to the slope of the PL tails of the density PDF. As a result, a second PL tail in $p(s)$ having a PL slope less than 1.5 can be seen in the $\epsff(s)$ curve as the range where the slope, $d\epsff(s)/ds$, is negative (see e.g., \autoref{fig:PDFEff_analytical}). In the absence of a second PL tail, we do not expect a change in slope at high $s$ in the $\epsff(s)$ curve. Since gas in the second PL tail is present in discs, we do not expect the second PL tail to be resolvable when the discs are unresolved. To test this, we repeat simulation v2M5 with five different maximum resolutions: 1611 AU, 805 AU, 402 AU, 201 AU, and 101 AU; the 201 AU case is the standard one that we use elsewhere in the paper. In \autoref{fig:ResdiscSize} (top), we show $\epsff(s)$ for these 5 simulations; the curves are averaged over the SFE range 4\% and 6\% for each individual simulation. 
The blue, red and green curves, which belong to simulations with a minimum cell size $\Delta x\approx 1611, 805$ and $402$ AU, respectively, are not able to resolve the discs, a failure that manifests as the absence of the characteristic local maximum and minimum that we have shown is associated with the development of the second PL tail. By contrast, there is a clear indication of a second PL tail in the orange and cyan curves, which correspond to resolutions of 201 and 101 AU, respectively. These simulations are able to resolve the discs and capture the gas dynamics more accurately. The disc size seen in this simulation has a radius $r_{\rm d} \sim$ 2000 AU. Therefore we estimate that capturing the disc / second PL transition requires that $\mathcal{H}\equiv r_{\rm d}/\Delta x \gtrsim10$. 

To check that this condition matches other simulations, and also whether the resolution requirement is time-dependent, we also repeat the v1M10 simulation at a resolution of $\Delta x = 201$ AU, a factor of 4 lower than our fiducial resolution of 50 AU for this case. The bottom panel of \autoref{fig:ResdiscSize} shows the resulting $\epsff(s)$ curves at SFEs of 5\% (averaged over SFEs of 4-6\%) and 13\% (averaged over 12-14\%). As expected, a minimum cell size of $\Delta x\approx 201$ AU is not sufficient to resolve the discs in this case. For $\Delta x\approx 50$~AU resolution, we begin to see a flattening of the $\epsff(s)$ curve at $s>5$, indicating the first hints of discs and the formation of a PL slope $\alphad$ close to the asymptotic value of 1.5. This is consistent with our resolution criterion of $\mathcal{H}\gtrsim 10$, because in this case we find disc radius $\sim 300$~AU. Thus our maximum resolution reaches $\mathcal{H}\approx 6$, close to but not quite reaching our requirement $\mathcal{H} \gtrsim 10$.

%This is expected (cite??)

% Don't change these lines
\bsp	% typesetting comment
\label{lastpage}
\end{document}